\RequirePackage{fix-cm}
\documentclass[smallextended]{svjour3}       
\smartqed  
\usepackage{graphicx}
\usepackage{xspace} 
\usepackage{bm}
\usepackage{xcolor}
\usepackage{amsmath}
\usepackage{amssymb}

\journalname{Journal of Low-temperature Physics}
\begin{document}

\title{London penetration depth measurements using tunnel diode resonators}

\author{Russell Giannetta \and Antony Carrington \and Ruslan Prozorov}

\institute{Russell Giannetta \at
Loomis Laboratory of Physics, 
University of Illinois at Urbana-Champaign,
1110 W Green Street, Urbana, Illinois 61801, USA         
           \and
Antony Carrington \at
H. H. Wills Physics Laboratory,
University of Bristol, Bristol, BS8 1TL England
              \and
Ruslan Prozorov \at
Ames Laboratory and Department of Physics \& Astronomy, Iowa State University, Ames, Iowa 50011,	USA
}

\date{Received: June 2021 / Accepted: date}

\maketitle

\begin{abstract}
The London penetration depth $\lambda$ is the basic length scale for electromagnetic behavior in a superconductor.  Precise measurements of $\lambda$ as a function of temperature, field and impurity scattering have been instrumental in revealing the nature of the order parameter and pairing interactions in a variety of superconductors discovered over the past decades.
Here we recount our development of the tunnel-diode resonator technique to measure $\lambda$ as function of temperature and field in small single crystal samples.  We discuss the principles and applications of this technique to study unconventional superconductivity in the copper oxides and other materials such as iron-based superconductors.  The technique has now been employed by several groups world-wide as a precision measurement tool for the exploration of new superconductors.

\keywords{London penetration depth \and tunnel-diode resonator}
\end{abstract}

\section{Introduction}

Since the discovery of superconductivity in the copper oxides \cite{htsc}, an immense number of new superconductors have been discovered, most with ground states far more complex than the elemental materials first addressed in BCS theory \cite{bcs}.   In this paper, we discuss some of our work using tunnel diode resonators to measure the London penetration depth in several newly discovered superconductors.  We review how penetration depth is used as a probe of the order parameter and then discuss details of our first tunnel diode resonator design.  We then discuss experiments to study phenomena unique to unconventional pairing states, multiband superconductivity, impurity effects and electronic phase transformations. 

\section{Order parameter and gap function}

The motivation for nearly all the measurements discussed here is to determine the superconducting order parameter, first defined in BCS theory as an anomalous average that grows continuously from zero at the transition temperature $T_{c}$\cite{bcs},

\begin{equation} 
	\left\langle {{c_{\bm k{\kern 1pt} \alpha }}{c_{ - \bm k{\kern 1pt} \beta }}} \right\rangle
\end{equation}

\noindent The $c_{\bm k\alpha}$ are fermion destruction operators, $\bm k$ is the wavevector and $\alpha$,$\beta$ are spin indices.  Physically measurable quantities are more directly related to a gap matrix, 
\begin{equation}
	{\Delta _{{\kern 1pt} \alpha {\kern 1pt} \beta }}\left( {\bm k} \right) = \sum\limits_{\bm k} {V\left( {\bm k,\bm k'} \right)} \left\langle {{c_{\bm k'{\kern 1pt} \alpha }}{c_{ - \bm k'{\kern 1pt} \beta }}} \right\rangle 
\end{equation}

\noindent $ {\Delta _{{\kern 1pt} \alpha {\kern 1pt} \beta }}\left( {\bm k} \right)$ is determined self-consistently through the gap equation where $V({\bm k},{\bm k}')$ is the pair interaction responsible for superconductivity.  Fermions may pair in singlet, triplet or in principle, admixtures of these two spin states.   The spin character of the order parameter is best determined by nuclear magnetic resonance.    Triplet superconductivity is very rare and the materials discussed here all pair in a spin singlet state.  In that case the gap function simplifies to \cite{sigrist}, 
\begin{equation}
	{\Delta _{{\kern 1pt} \alpha {\kern 1pt} \beta }}\left( {\bm k} \right) = i\,{\sigma _y}\Delta \left( {\bm k,T} \right)
\end{equation}

\noindent where $\sigma_{y}$ is a Pauli matrix and $\Delta$ will be referred to as the gap function.  In the simplest version of BCS theory, $\Delta$ depends on temperature but is independent of $\bm k$, a case known as isotropic $s$-wave pairing.  However, in general $\Delta$ depends on $\bm k$. The requirement that the pair wavefunction be antisymmetric under interchange of fermions means that the gap function for a singlet state must have even parity ($s$, $d$,...).   

\indent The discovery of superconductivity in the copper oxides prompted several authors to propose a pair interaction based on the exchange of antiferromagnetic spin fluctuations \cite{scalapino,annett}.  For these materials, this mechanism favors a $d_{{x^2} - {y^2}}$ gap function,

\begin{equation}
	\Delta \left( {{d_{{x^2} - {y^2}}}} \right) = {\Psi_d}\left( T \right)\left( \cos(k_{x}a)-\cos(k_{y}a) \right)
	\label{eqdgap}
\end{equation}

\begin{figure}[tbh]
	\centering
	\includegraphics[width=12cm]{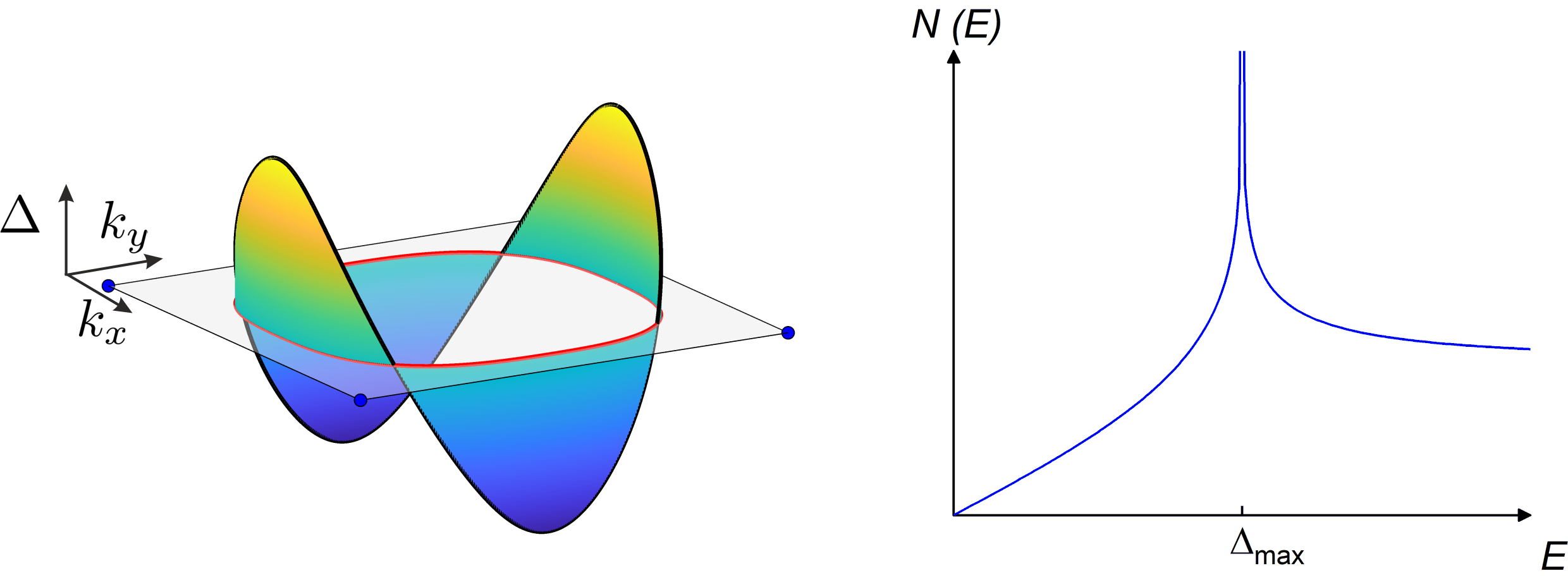}%
	\caption{(Color online) $d$-wave gap function suggested for pairing state of copper oxide superconductors shown here in the $ab-$plane of a cylindrical Fermi surface. Also shown the total (integral) density of states, $N(E)$. The characteristic linear behavior due to nodal regions results in a $T-$linear low-temperature $\lambda(T)$.}
	\label{fig1}
\end{figure}

\noindent $\Psi_{d}(T)$ is the magnitude of the gap whose temperature dependence will be discussed momentarily.  Alternatively, a different representation $\Delta \left( {{d_{{x^2} - {y^2}}}} \right) = {\Psi_d} \cos 2\phi$, (where $\phi$ is the in-plane azimuthal angle) is often used which is similar to Eq.\ \ref{eqdgap} for a circular Fermi-surface and has the same symmetry.  Figure \ref{fig1} shows the $d_{{x^2} - {y^2}}$ gap on a cylindrical Fermi surface.  This particular state has nodes at 45$^\circ$ relative to the crystalline axes while a $d_{xy}$ state would have nodes along the axes.  The existence of a sign-changing gap function can rule out some pairing mechanisms and favor others so its identification inspired an intense world-wide effort.

Nodes have important experimental consequences since they determine the low temperature behavior of quantities such as specific heat \cite{moler} and penetration depth \cite{hardy}. This arises through the thermal population of quasiparticles, whose occupation number,

\begin{equation}
	f\left( {{E_{\bm k}}} \right) = \left( {1 + {e^{ - {{{E_{\bm k}}} \mathord{\left/
						{\vphantom {{{E_{\bm k}}} {{k_B}T}}} \right.
						\kern-\nulldelimiterspace} {{k_B}T}}}}} \right)^{-1}\
\end{equation}

\noindent involves the excitation energy, 

\begin{equation} 
	E_{\bm k} = \sqrt {{\varepsilon _{\bm k}^2} + {{\left| {\Delta \left( {\bm k} \right)\,} \right|}^2}} 
\end{equation}

\begin{figure}[tbh]
	\centering 
	\includegraphics[width=8.5cm]{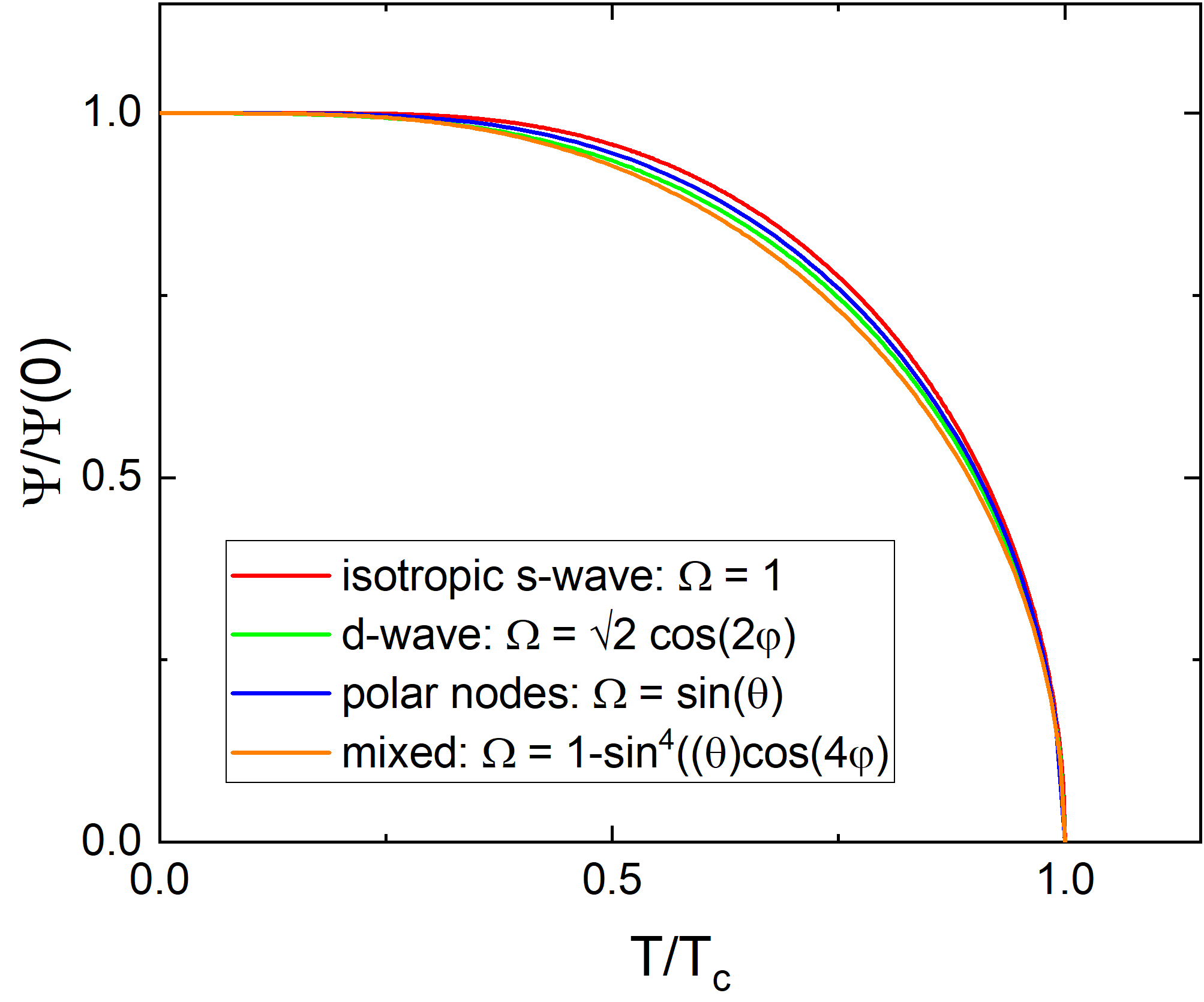} 
	\caption{(Color online) Temperature-dependence of the normalized gap amplitude defined in Eq.\ref{Del}. $\theta$ and $\phi$ are polar and azimuthal angles of $\bm k$.  Different $\Omega (\bm{k})$ functions produce only small differences in $\Psi(T) / \Psi(0)$ at intermediate temperatures.} 
	\label{fig2} 
\end{figure}

\noindent $\epsilon_{\bm k}$ is the band energy measured relative to the chemical potential.  The right panel of figure \ref{fig1} shows the density of quasiparticle states $N(E)$ for the $d_{{x^2} - {y^2}}$ gap.  For a gap without nodes there are no states below $\Delta_{max}$.  Nodes lead to a linear contribution $N(E) \sim {E}$.  As shown in figure \ref{fig2} , below about $0.25 {\ } T_{c}$ the gap function is nearly temperature-independent. The low temperature variation of the penetration depth then comes entirely from the variation of the gap in $\bm k$-space.  The linear quasiparticle density of states leads to a power law temperature dependence while a finite gap everywhere on the Fermi surface leads to an exponentially-activated dependence. Low temperature measurements with sufficient precision can therefore indicate the presence of nodes, or at least place a lower limit on magnitude of the gap. It would seem that experiments relying solely on quasiparticle occupation cannot distinguish a state with nodes but no sign change from the sign-changing $d_{{x^2} - {y^2}}$ state.  However, we will see that the sign change of the $d$-wave gap function, though more directly accessible through Josephson and flux quantization measurements \cite{dale,tsuei}, can reveal itself through penetration depth measurements at low temperatures. Later, in our discussion of multiband superconductors, we discuss how $\lambda$ measurements together with controlled impurity scattering can also lead to phase sensitivity.

Many of our measurements focus on the region well below $T_{c}$ where the temperature dependence of the gap function can be ignored, regardless of its momentum dependence.  Within the weak-coupling (BCS) approximation the pair interaction is assumed to have the form, $V(\bm{k},\bm{k}^{\prime})=V_{0}\Omega(\bm{k})\Omega(\bm{k}^{\prime})$.  This leads to a gap function of the general form,
\begin{eqnarray}
	\Delta(T,\bm{k})=\Psi(T)\Omega(\bm{k})\,.
	\label{Del}
\end{eqnarray}

\noindent where $\langle\Omega^{2}({\bm k})\rangle=1\,,$ the average taken over the Fermi surface. $\Psi\left(0\right)$ is obtained by solving the following self-consistent equation \cite{KP2021PRB},
\begin{equation}
	-\ln\frac{T}{T_{c}}
	=\sum_{n=0}^{\infty}\left(\frac{1}{n+1/2}-\left\langle \frac{2\pi T\Omega^{2}(\bm k)}{\sqrt{\left(\pi T\left(2n+1\right)\right)^{2}+\Delta^{2}(T,\bm k)}}\right\rangle \right)	
	\label{selfcon}
\end{equation}

\noindent The brackets denote a Fermi surface average and $\hslash\omega_{n}=\pi T\left(2n+1\right)$ are Matsubara frequencies.  A very good approximation for the temperature dependence of the gap function is given by \cite{KP2021PRB},
\begin{equation}
	\frac{\Psi(t)}{\Psi(0)}=\tanh\left(e^{\gamma}\sqrt{\frac{8(1-t)}{7\zeta(3)\,t}}\frac{e^{\langle\Omega^{2}(\bm k)\ln|\Omega(\bm k)|\rangle}}{\sqrt{\langle\Omega^{4}(\bm k)\rangle}}\right).\qquad
	\label{Einzel}
\end{equation}

\noindent  where $\gamma$ = $0.57721$ is Euler's constant and $\zeta (3)$ = $1.202$ is the Riemann zeta function.  Figure \ref{fig2} shows that the ratio in equation \ref{Einzel} is a nearly universal function of $t{\ }={\ }T/T_{c}$ and only weakly dependent on the $\bm k$-variation of the gap function.  Furthermore figure \ref{fig2} shows that $\Delta$ is nearly temperature independent below $0.25$  $T_{c}$, regardless of the $\bm k$ dependence of the gap.  
The approximately universal temperature dependence does not hold for multiband superconductivity where different gap functions exist on different Fermi surface sheets.  

\section{Penetration depth and superfluid density}

The basic experimental feature of a superconductor is perfect diamagnetism, \textit{i.e.}, the Meissner effect. Using the second London equation and Ampere’s law 
the magnetic field in a superconductor obeys,

\begin{equation}
	\nabla^{2}\mathbf{B}=\lambda^{-2}\mathbf{B}
	\label{london1}
\end{equation}

\noindent where $\lambda$ is the London penetration depth.   Equation \ref{london1} implies that an applied magnetic field decays exponentially into a superconductor on the length scale of $\lambda$.  More generally, the penetration depth is anisotropic.   The second London equation then relates the supercurrent components to the vector potential $A_k$ through a superfluid density tensor $\lambda_{ik}^{-2}$,  
\begin{equation}
	\mu_{0}j_{i}=-\lambda_{ik}^{-2}A_{k}
	\label{london2}
\end{equation}

\noindent where we assume the London gauge, $\nabla\cdot {\bm A}=0$. The measured penetration depth components $\lambda_{i}$ are then obtained from the principal axis components of $\lambda_{ik}^{-2}$.   $\lambda_{i}$ therefore refers to the penetration depth for the current $j_i$, {\it not} the field component $B_i$.  In layered superconductors such as copper oxides, $\lambda_{a,b}$ , the penetration depth for currents within the conducting planes is quite different from $\lambda_c$, the penetration depth for interplane currents.   Unless otherwise noted we will denote the in-plane penetration depth as simply $\lambda$. 

Microscopic treatments of the penetration depth can be found in many excellent texts and articles.  For the present we recall a useful semiclassical result which holds in London limit where the current is locally proportional to the vector potential \cite{gross},

\begin{equation}
\lambda^{-2} = \frac{\mu_{0} e^2}{\pi h^ 2} \int_{\rm{FS}} d\bm{S}\frac{v_i^2}{|v|} \left( 1 + 2\int_{\Delta(\bm{k})}^\infty \frac{{\rm{d}}f}{{\rm{d}}E} \frac{E{\rm{d}}E}{\sqrt{E^2 - \Delta_{\bm{k}}^2}} \right)  = \frac{\mu_{0} e^2}{m} n_{S}(T) 
\label{sfd}
\end{equation}

\noindent  The $v_i$ factors are Fermi velocities along the $i \in (x,y,z)$ direction, while the gap function and quasiparticle energy were introduced earlier.  Equation \ref{sfd} consists of two competing terms. The temperature-independent term comes from a diamagnetic current that is present in both normal and superconducting states when a magnetic field is applied. It is proportional to the full density of carriers $n_0$ and does not depend on the superconducting parameters. The second term comes from a paramagnetic current that depends on both the gap function and the temperature \cite{tinkham}.  This term defines the normal fluid density $n_{N}$.  The superfluid density is given by the difference $n_{S}(T)$= ${n_0}$ - $n_{N}(T)$.  Above $T_{c}$ , the two currents cancel, $n_S$ vanishes and the penetration depth is infinite.  In liquid helium physics one directly measures the superfluid density but in superconductivity it is $\lambda$ that is more often measured.  

\indent Figure 2 shows that below approximately 0.25 $T_c$ the energy gap for $s$ and $d$ wave pairing is essentially temperature independent.  In this region we can use equation \ref{sfd} to find the temperature dependence of the normalized superfluid density (also known as the phase stiffness).  For $s$-wave pairing we recover the standard BCS result, 

\begin{equation}
\rho_{S}=\frac {n_{S}}{n_{0}}=\left(\frac{\lambda(0)}{\lambda(T)}\right)^2 =1 - \sqrt{\frac{2\pi\Delta_s(0)}{T}} \exp\left(\frac{-\Delta_s(0)}{k_BT}\right)
	\label{rhos}
\end{equation}

\noindent  For a $d$-wave gap function, $\Delta_d(T)=\Delta_0\cos{(2\phi)}$ where $\Delta_0=2.14T_c$. Assuming an isotropic or spheroidal Fermi surface, $\rho_{S}$ is linear at low temperatures \cite{prozorov1},

\begin{equation}
	\rho_{S} = \,1 - \frac{{2\,\ln 2}}{{{\Delta _d(0)}}}T.
	\label{dwave}
\end{equation}

\noindent More generally, for a superconductor with $N$ line nodes,

\begin{equation}
\rho_{s}=1-\frac{N\ln2}{\left.d\left|\Delta\left(\varphi\right)\right|/d\varphi\right|_{\textrm{node}}}T
	\label{gendwave}
\end{equation}

\noindent where the denominator is the slope of the gap angular dependence at the node \cite{Xu95}. \indent   The predicted linear temperature dependence arising from nodal quasiparticles was first measured in optimally doped YBa$_2$Cu$_3$O$_{6+x}$ (Y123) crystals by Hardy \textit{et. al.} \cite{hardy} using a superconducting microwave resonator.  This important experiment was crucial to the wide acceptance of nodes in the gap function of copper oxide superconductors and gave support to the hypothesis of $d$-wave pairing in these materials. The temperature dependence is crucial since the in-plane magnetic response (equation 12) is isotropic for a $d$-wave superconductor with tetragonal lattice symmetry.  Therefore, changing the direction of the magnetic field cannot distinguish $d$-wave from $s$-wave. 

\indent For an arbitrary angular variation of the gap, $\Omega({\bm k})$, the amplitude $\Psi(T)$ is found by solving equation 8 at zero temperature, after which the nodal derivative $d\Delta/d\varphi=\Psi(0)d\Omega/d\varphi$ is determined. The shape of the Fermi surface affects $d\lambda/dT$ through its dependence on $\langle v^2/|v|\rangle_{\rm{FS}}$ at the node, as indicated in equation 12. The nodes may also be inequivalent, in which case separate derivatives must be calculated.

\indent The issue of strong versus weak coupling sometimes arises in the interpretation of data.  Strong coupling increases the magnitude of the gap relative to $T_c$ so it is more likely to extend the low temperature region over which power laws are determined by the $\bm k$ dependence of the gap function.   For a $d$-wave superconductor, the coefficient of $T$ in equation \ref{dwave} is affected by the magnitude of gap function but it is also affected by its specific shape in $\bm k$ space and the details of the Fermi surface, so strong coupling effects are difficult to isolate.  Kogan and Prozorov \cite{KP2021PRB} have shown that claims of strong-coupling obtained by fitting to an exponential dependence of $\lambda$ can often be erroneous.  Later, we discuss two band superconductors.   There, weak coupling implies a generalization of the familiar $N(0)V \ll 1$ , $V$ = $const$ criterion \cite{tinkham} to a $2 \times 2$ matrix $N_{i}(0) V_{ij}$ that involves the pair interaction both within and between bands.  

\section{Nonlinear Meissner Effect}

The development of our technique for measuring penetration depth was motivated by a theoretical paper by S.\ Yip and J.\ Sauls \cite{ys} which suggested that penetration depth measurements as a function of magnetic field could provide a stringent test of $d$-wave pairing and potentially locate the directions of the nodes of the gap function.  They termed this phenomenon the nonlinear Meissner effect and predicted that it would be observable at low temperatures in sufficiently pure Y123 crystals.   For a simple picture, recall that the energy of a quasiparticle in the BCS state depends on the direction of its momentum relative to the net superfluid velocity $\bm v_{S}$ ,    
\begin{equation}
	{E_{\bm k}} = {E_{\bm k}}\left( {{{\bm v}_S} = 0} \right) + \hbar {\kern 1pt} \bm k \cdot {\bm v_S} 
\label{doppler}
\end{equation}

\noindent Quasiparticles with momentum parallel to $\bm v_{S}$ have higher energy than those antiparallel, i.e., a Doppler shift.   The counter-moving excitations reduce the net supercurrent and therefore increase the penetration depth.   The penetration depth is also increased by the field in a fully-gapped superconductor but the effect is small, the correction being of order $(H/H_0)^2$ where $H_0$ is a scaling field of order the thermodynamic critical field. In addition, the correction vanishes exponentially with temperature due to the presence of a finite energy gap everywhere on the Fermi surface.   

\indent Yip and Sauls showed that $d$-wave pairing changes the picture entirely.    Due to the presence of nodes, counter-moving quasiparticles in the $d$-wave state are more likely to be populated by the application of a field. This occurs at fields low enough that suppression of the gap itself by the field is small.  The repopulation  reduces the supercurrent and leads to an increase of the penetration depth that is {\it linear} in the applied field. At finite temperature this correction is reduced at low field ($H<H_{0}(T)$) so the total change with field grows as the temperature is reduced.  Their prediction at $T$ = $0$ was,

\begin{equation}
		\lambda \left( {H} \right) = \lambda \left( {H = 0} \right)\left( {1 + \frac{H}{{{H_0}}}} \right)
\end{equation}

\noindent where $H_0$ is predicted to be $\sqrt{2}$ larger for $H\perp$ node compared to $H\|$ node.
\indent  The observation of a nonlinear Meissner effect posed a significant experimental challenge.  The need to work in the Meissner state requires $H < H_{c1}$ (about 100~Oe for $H\| ab$ plane of Y123 ) while $H_{0} \sim 2$\,T in Y123.   The largest predicted field-dependent correction to $\lambda$ is less than 1 percent.  Since $\lambda(0) \sim$ 1500\,\AA, it was necessary to resolve Angstrom-level changes in $\lambda$ using crystals whose maximum dimension was typically less than 1\,mm.   

\indent The need to apply a DC magnetic field in addition to a tiny RF probing field ruled out a superconducting microwave resonator approach.  One of us (RWG) recalled a Cornell low temperature physics group seminar given in the late 1970’s by Craig Van Degrift.  He had built an extremely stable tunnel diode resonator (TDR) to measure tiny changes in the dielectric constant of liquid helium \cite{vDG}.  The device operated at a few MHz using a copper coil and a capacitor for the resonant cavity.   The high sensitivity and absence of superconducting components suggested that the TDR would be an ideal device to search for the nonlinear Meissner effect.  

\section{Measurement of Penetration Depth}

Using the London equation it is a standard exercise to show that the induced magnetic moment $m$ of a superconducting slab in a uniform magnetic field $H$ is given by,

\begin{equation}
	m = Lwd\left( {1 - \frac{{2{\kern 1pt} \lambda }}{d}{\kern 1pt} \tanh \frac{d}{{2{\kern 1pt} \lambda }}} \right)H
	\label{london}
\end{equation}

\noindent Here $L$, $w \gg d$ where $L$, $w$ and $d$ are the length, width and thickness of the slab.  The field is applied parallel to the longest dimension therefore minimizing demagnetizing effects.   In our measurements $H$ is an RF  field. Unfortunately, a thin slab, a sphere and a cylinder are the only shapes for which the London equation can be solved exactly.  The irregular shapes encountered in experiments can usually be treated with the following approximate expression,

\begin{equation}
	m = \frac{{Vol}}{{1 - N}}\left( {1 - \frac{\lambda }{R}\tanh \left( {{\lambda  \mathord{\left/
					{\vphantom {\lambda  R}} \right.
					\kern-\nulldelimiterspace} R}} \right){\kern 1pt} } \right)H\quad 
	\label{R}
\end{equation}

\noindent where $N$ is a demagnetizing factor and $R$ is an effective sample dimension for which Prozorov \textit{et al.} derived an approximate analytical expression \cite{prozorov1,prozorov2}.  If the sample is now inserted into the coil of an $LC$ oscillator, the oscillation frequency increases by an amount,
\begin{equation}
	\delta f = f(with\,sample) - f(without\;sample) = G\,m
	\label{pullout}
\end{equation}

\noindent where $G$ depends on the effective volume of the coil and the oscillation frequency.  While the prefactor $G{\ } Vol/(1-N)$ is difficult to calculate reliably, it can be directly measured by removing the sample from the coil {\it in situ} at the base temperature where the $\lambda$-dependent term is typically $10 ^{-3}$ or smaller.  For typical sample sizes, the frequency shift in equation \ref{pullout} ranges from 1 $\sim $ 100 kHz and repeats to within 1-2 ppm.   Once $G{\ } Vol/(1-N)$ has been obtained, we measure the change in resonator frequency as the temperature or an external DC magnetic field is varied. A more complete account of the calibration, including updated formulas for the effective dimensions can be found in ref \cite{Prozorov2021}. 

The small size of the $\lambda$-dependent term in $m$ means that one cannot simply insert the sample into the coil, measure the change in oscillation frequency and determine $\lambda$.  To do that would require knowing the sample geometry, field orientation and then numerically solving the London equation for $m$,  all with far more precision than is currently feasible.  What can be done is to numerically calculate the effective sample dimension $R$ in equation \ref{R}, measure the change in frequency as the sample temperature changes and convert that change into  $\Delta \lambda$ = $\lambda(T)-\lambda(0)$ to a precision of a few percent.  The normalized superfluid density is then given by,

\begin{equation}
	{\rho _{{\kern 1pt} S}} = {\left( {\frac{{\lambda \left( 0 \right)}}{{\lambda \left( T \right)}}} \right)^2} = \frac{1}{{{{\left( {1 + {{\Delta \lambda } \mathord{\left/
								{\vphantom {{\Delta \lambda } {\lambda \left( 0 \right)}}} \right.
								\kern-\nulldelimiterspace} {\lambda \left( 0 \right)}}} \right)}^2}}}
\end{equation}

\noindent The zero temperature penetration depth $\lambda(0)$ is often obtained from other techniques such as $\mu$SR, small angle neutron scattering or infrared reflectively.  These techniques have far less sensitivity to $\Delta \lambda$ but can determine $\lambda(0)$ to reasonable precision.  Later, we will discuss an aluminum-plating technique which allows us to measure the full value of $\lambda$ with the tunnel diode resonator alone.  

Commercial magnetometers lack the sensitivity for the measurements discussed here.   For example, a perfectly diamagnetic cylinder 1 mm in diameter and 0.1 mm thick (slightly larger than our typical sample) has a moment $m_{0}\approx-6\times 10^{-6}$\,emu (1 emu{\ }={\ }erg/G).  A change in penetration depth $\Delta\lambda\left(T\right)$ for this sample corresponds to a change in magnetic moment $\varDelta m\approx2m_{0}\Delta\lambda/R$.  For $\Delta\lambda=$ ~1 \AA\, this corresponds to $\varDelta m\approx2m_{0}\Delta\lambda/R\approx3\times10^{-12}$\,emu, approximately 4 orders of magnitude smaller than the sensitivity of commercial magnetometers. However, the required sensitivity can be attained with a tunnel diode resonator.

\section{Tunnel Diode Resonator}

\begin{figure}[tbh]
	\centering
	\includegraphics[width = 8.5cm]{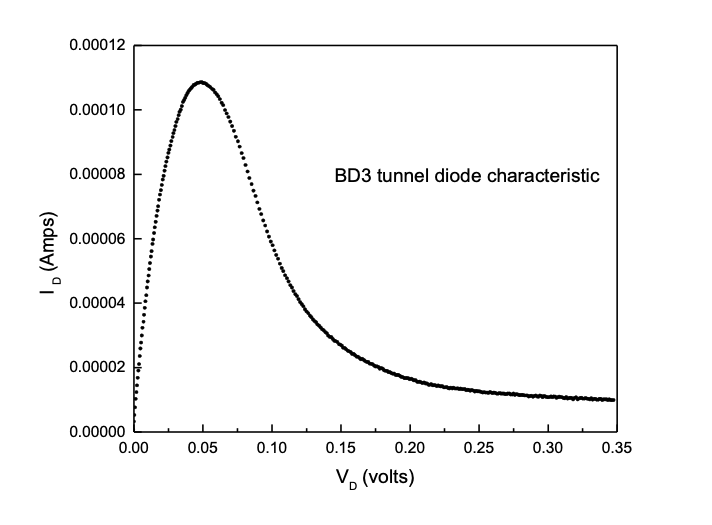}
	\caption{(Color online) Current voltage characteristic of BD3 tunnel diode.}
	\label{fig3}
\end{figure}

\begin{figure}[tbh]
	\centering
	\includegraphics[width = 8.5cm]{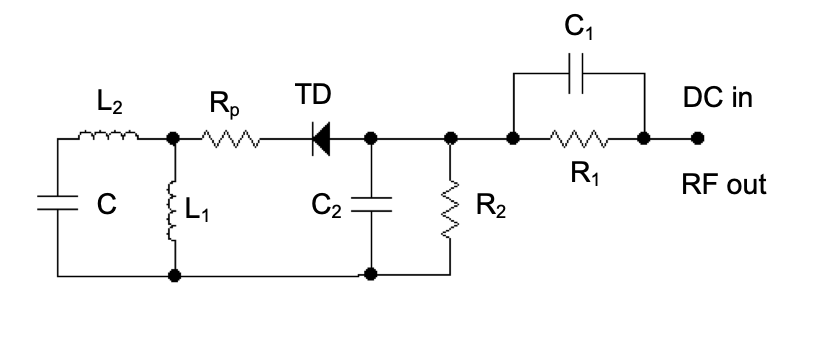}
	\caption{(Color online) Low temperature oscillator circuit schematic diagram. }
	\label{fig4}
\end{figure}

Tunnel diode resonators have been used in low temperature physics since the 1960's \cite{heybey,boghosian,tedrow}. Their advantages include simplicity, high sensitivity, low power dissipation and the ability to operate in static magnetic fields of many Tesla. Figure \ref{fig3} shows the IV characteristic of a BD3 tunnel diode. At the inflection point the diode has its peak negative differential resistance , $-R_{D} $, which will lead to sustained oscillations when incorporated into the circuit shown in Figure \ref{fig4}.  $L_{2}$ is the sample coil and $L_{1}$ is a tap coil that adjusts the impedance of the tank circuit, making the oscillator `marginal' which reduces substantially the frequency noise level\cite{vDG}.  $R_{P}$ is used to reduce parasitic resonances. The other resistors maintain the correct DC bias while $C_{1}$ and $C_{2}$ couple the RF signal out of the oscillator and provide an AC ground, respectively.  At resonance, the tapped $LC$ tank circuit presents a resistive impedance ${\eta ^2}{Q ^2}r$ where $Q$ is the quality factor, $r$ is the coil AC resistance and ${\eta}={L_1}/(L_{1}+L_{2})$ is the tap fraction.  The resonant frequency $f = 1/{2 \pi}(C(L_{1}+L_{2}))^{1/2}$ is typically 10-15 MHz.  Typically, oscillations do not begin until well below 77\,K when the onset condition $R_{D}-R_{P} = {\eta ^2}{Q ^2}r$ is met.  There are many small corrections to the frequency that depend on $Q$ and the other circuit component values but these are independent of the superconducting sample properties and are therefore constant throughout the measurement. Dissipation in the sample can affect $Q$ and therefore the oscillation frequency but this effect is negligible except very close to  $T_{c}$.   

The sub-mm dimensions of many superconducting crystals dictate the size of the sample coil $L_{2}$.  It must be small enough to be measurably perturbed by Angstrom-level changes in penetration depth yet large enough to generate a homogeneous RF magnetic field.  Ours are typically 1 cm long and 2 mm in diameter with a quality factor $Q$ = 100-200.  The coil is surrounded by a copper tube which makes the field more homogeneous, lowers its amplitude and provides temperature stability. The tunnel diode circuitry sits inside a shielded enclosure whose temperature is controlled near 2 K to within a few $\mu$K.   

\begin{figure}[tbh]
	\centering
	\includegraphics[width = 3.5cm]{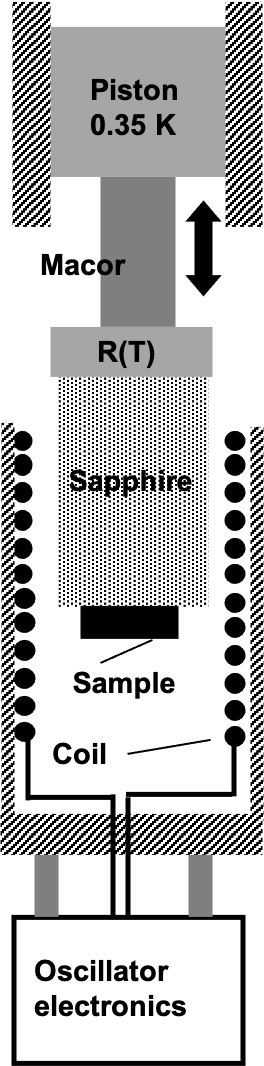}
	\caption{(Color online) Schematic of apparatus to measure penetration depth. Dimensions are not to scale. }
	\label{fig5}
\end{figure}

Figure \ref{fig5} shows a schematic of a typical apparatus \cite{carrington,organic}.  The sample is attached to a sapphire rod that can be moved in and out of the tunnel diode oscillator coil. The thermometer and heater stage, R(T), is located at the opposite end of the rod, outside the RF field of the coil.  The thermometry stage is in turn attached to a low thermal conductivity ceramic rod which is attached to a copper piston that slides inside a closely fitting copper cylinder.  The cylinder and piston are heat sunk to the $^3$He evaporator.  This scheme allows the sample temperature to be controlled anywhere from 0.35 – 150\,K, independent of tunnel diode circuit.  The latter is independently controlled near 2\,K but mechanically attached to the cylinder/piston assembly through a graphite tube. DC magnetic fields both parallel and perpendicular to the axis of the sapphire rod are applied with small superconducting coils inside the vacuum can.  The ability to control the oscillator temperature independently of the sample temperature and to remove the sample from the RF field {\it in situ} have proven to be indispensable for both calibration purposes and for eliminating frequency shifts due to applied magnetic fields. Variations on this design have been used with a dilution refrigerator to achieve sample temperatures below 50 mK \cite{bonalde,bonalde2,fletcher2007,fletcher2009,tanatar,wilcox}.

\begin{figure}[tbh] 
	\centering
	\includegraphics[width = 8.5cm]{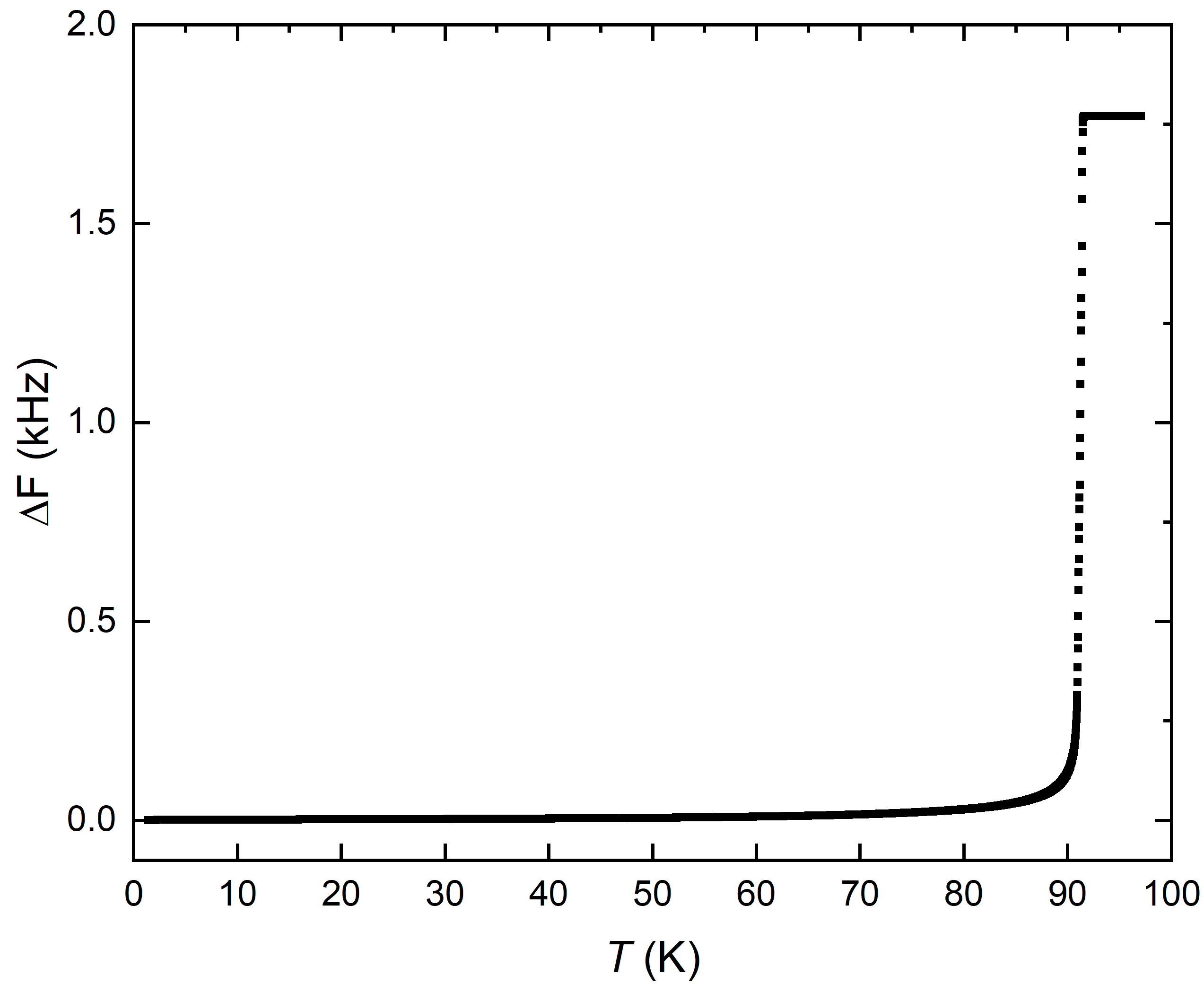}
	\caption{Frequency shift of the oscillator from low temperature to above $T>T_c$ for an optimally doped single crystal of YBa$_2$Cu$_3$O$_{6+x}$ with $H\|ab$. The low temperature portion of the graph is shown in Fig.\ 7.}
	\label{fig6}
\end{figure}

\indent Figure \ref{fig6} is an example of resonator data for a optimally doped single crystal of YBa$_2$Cu$_3$O$_{6+x}$ with $H\|ab$ and size $a\times b\times c = 0.8 \times 0.51\times 0.011 \rm{mm}^3$.  Above $T_{c}$ the curve is nearly flat due to the skin depth $\delta$ being larger than the sample thickness.  The sharp transition at $T_{c}$ is a useful check on the thermometry and sample homogeneity but it tells us little about the superconducting state beyond the existence of a Meissner effect.  With the vertical scale shown, the data for all superconductors looks very similar.  To distinguish an exponential from a power law temperature dependence we must work at temperatures well below $T_{c}$ and with the vertical scale expanded by 100-1000 depending on the sample size.

\section{Search for Nonlinear Meissner Effect}

\begin{figure}[tbh]
	\centering
	\includegraphics[width = 8.5cm]{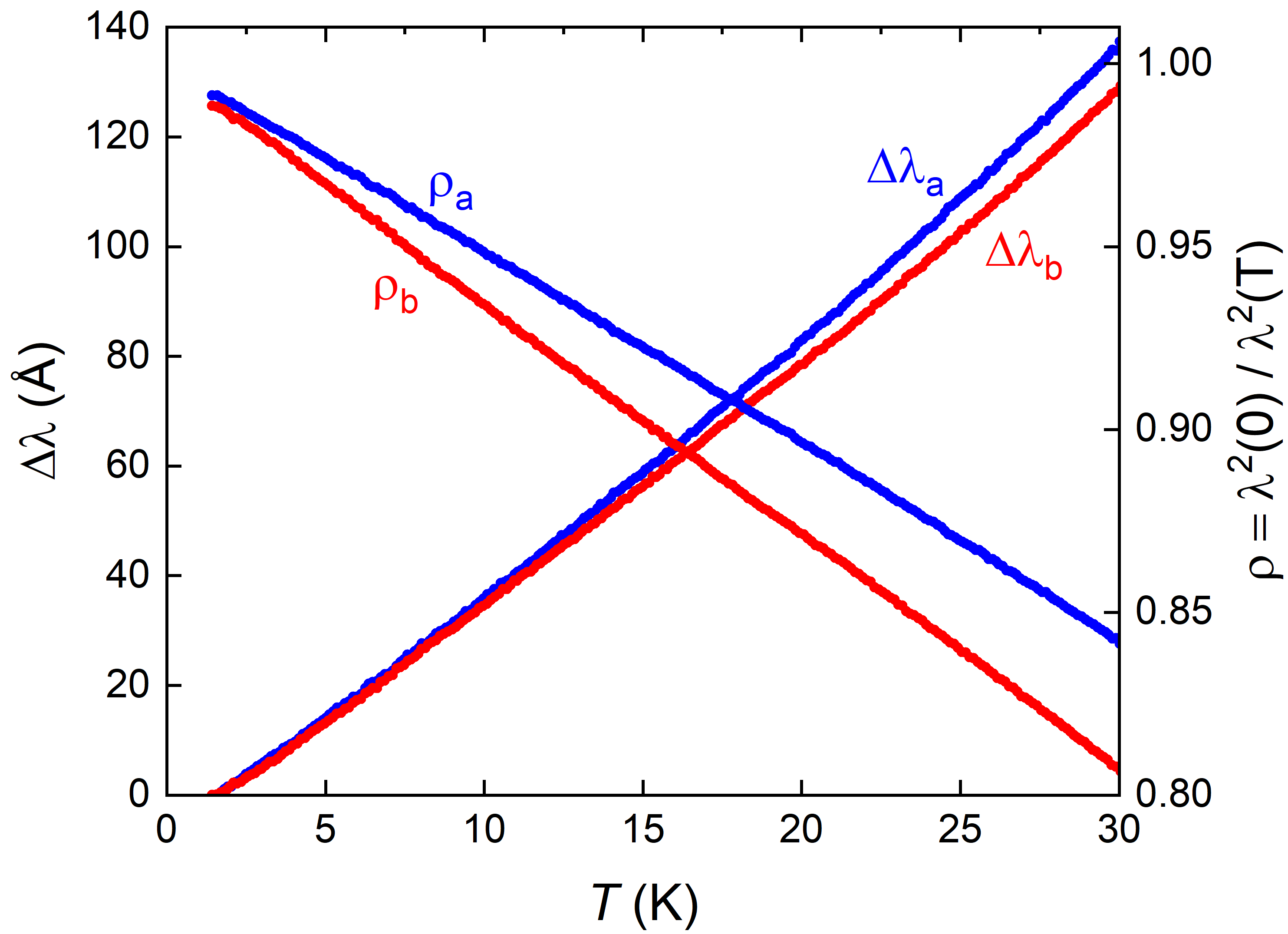}
	\caption{(Color online) Temperature dependence of the in-plane penetration depth ($\Delta \lambda$) of an optimally doped  single crystal of YBa$_2$Cu$_3$O$_{6+x}$  along the $a$ and $b$ axes.  The total frequency shift of the oscillator up to 30\,K is $\sim$ 3\,Hz. The normalised superfluid density ($\rho=\lambda^2(0)/\lambda^2(T)$) calculated with $\lambda_a(0)=1600$\,\AA~ and $\lambda_a(0)=1200$\,\AA~ is also shown.}
	\label{fig7}
\end{figure}

In the first application of the apparatus, Carrington \textit{et. al.} \cite {carrington} measured the temperature dependence of the penetration depth in a Y123 crystal.   Figure \ref{fig7} shows $\Delta \lambda_{a,b} (T)$ and $\rho_{a,b}(T)$ for supercurrents along the $a$ and $b$ directions in the copper-oxide plane.  We reiterate that $\Delta \lambda_{a,b}(T)$ are the directly measured quantities while the $\rho_{a,b}(T)$ use reported values of $\lambda_{a,b}(0)$ as inputs.  While all four quantities vary linearly at the lowest temperatures, the superfluid density is linear over a wider range than $\Delta \lambda$, as expected.  Our values of $d\rho_{a,b}/dT$ were within a few percent of those obtained by Hardy \textit{et al.} \cite{hardy}.  

\begin{figure}[tbh]
	\centering
	\includegraphics[width = 8.5cm]{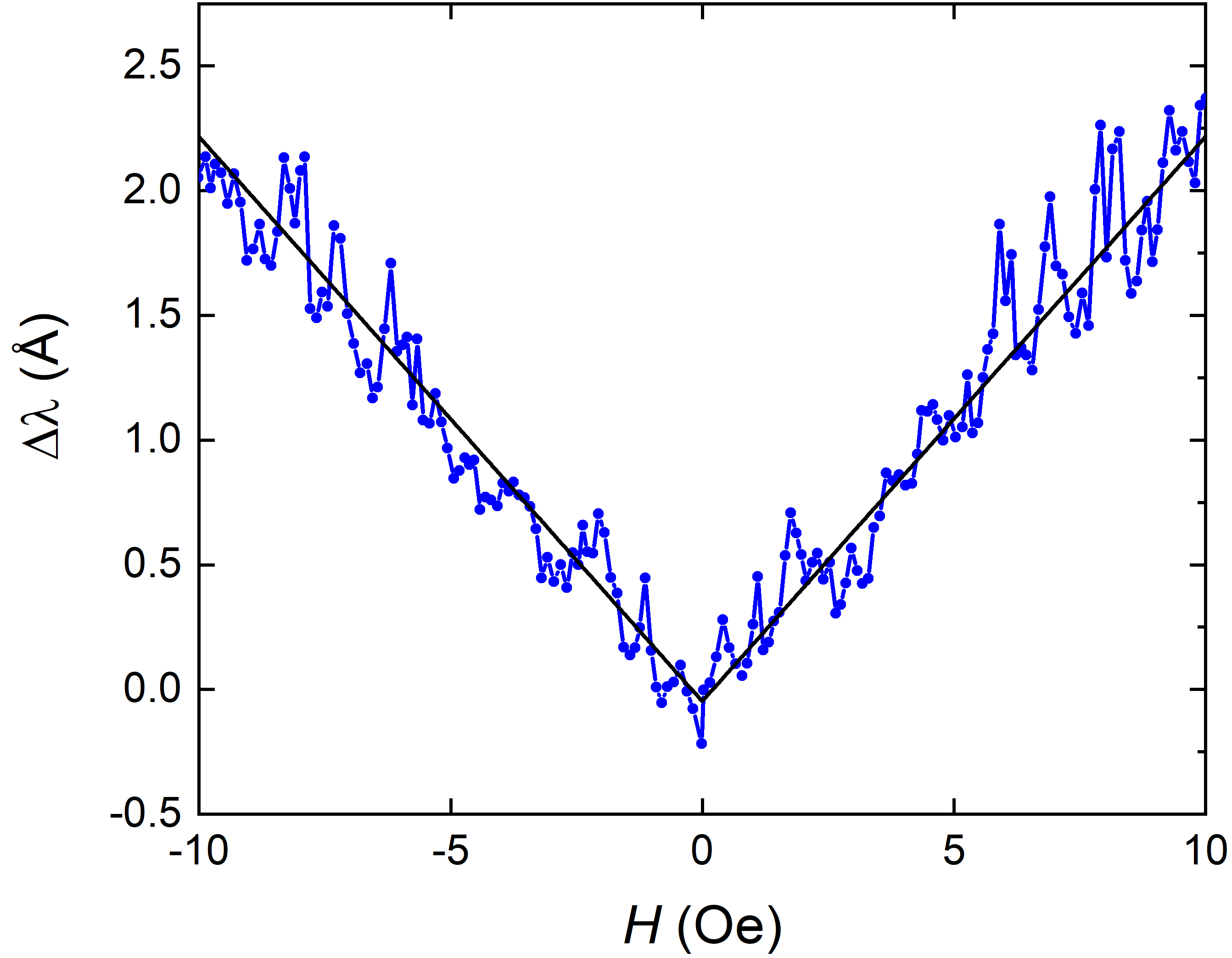}
	\caption{(Color online) Field dependence of the penetration depth for an optimally doped crystal of YBa$_2$Cu$_3$O$_{6+x}$ at $T=1.45$\,K, with both RF and DC fields parallel to $c$.  The applied field is shown, and the demagnetising factor $1/(1-N)\simeq 36$. The straight line is a linear fit. }
	\label{fig8}
\end{figure}

Our goal was to measure the magnetic field dependence of $\lambda$ and this data is shown in figure \ref{fig8}.  While we routinely observed a linear $H$ dependence, the data lacked other characteristics we were looking for.  Y123 is a type II superconductor so it admits quantized vortices.  Despite applying fields below the accepted lower critical field, it appears that vortices can still enter the sample and generate a field dependence to $\lambda$, possibly at the corners of the sample where the demagnetising effects are higher. In the mixed state, field dependent changes in $\lambda$ occur because the RF-field imparts a Lorentz force to a vortex.  The oscillating vortex carries magnetic field into the superconductor and therefore increases the total penetration depth.  This vortex contribution, known as the Campbell penetration depth, adds in quadrature with the desired contribution from the Meissner state:\cite{campbell1,campbell2,prozorov3,wu,tea},   

\begin{equation}
\lambda ^2 = \lambda^2_{\rm{Meissner}}+\lambda^2_{\rm{Campbell}}
\label{campbell}
\end{equation}

\begin{figure}[tbh]
	\centering
	\includegraphics[width = 8.5cm]{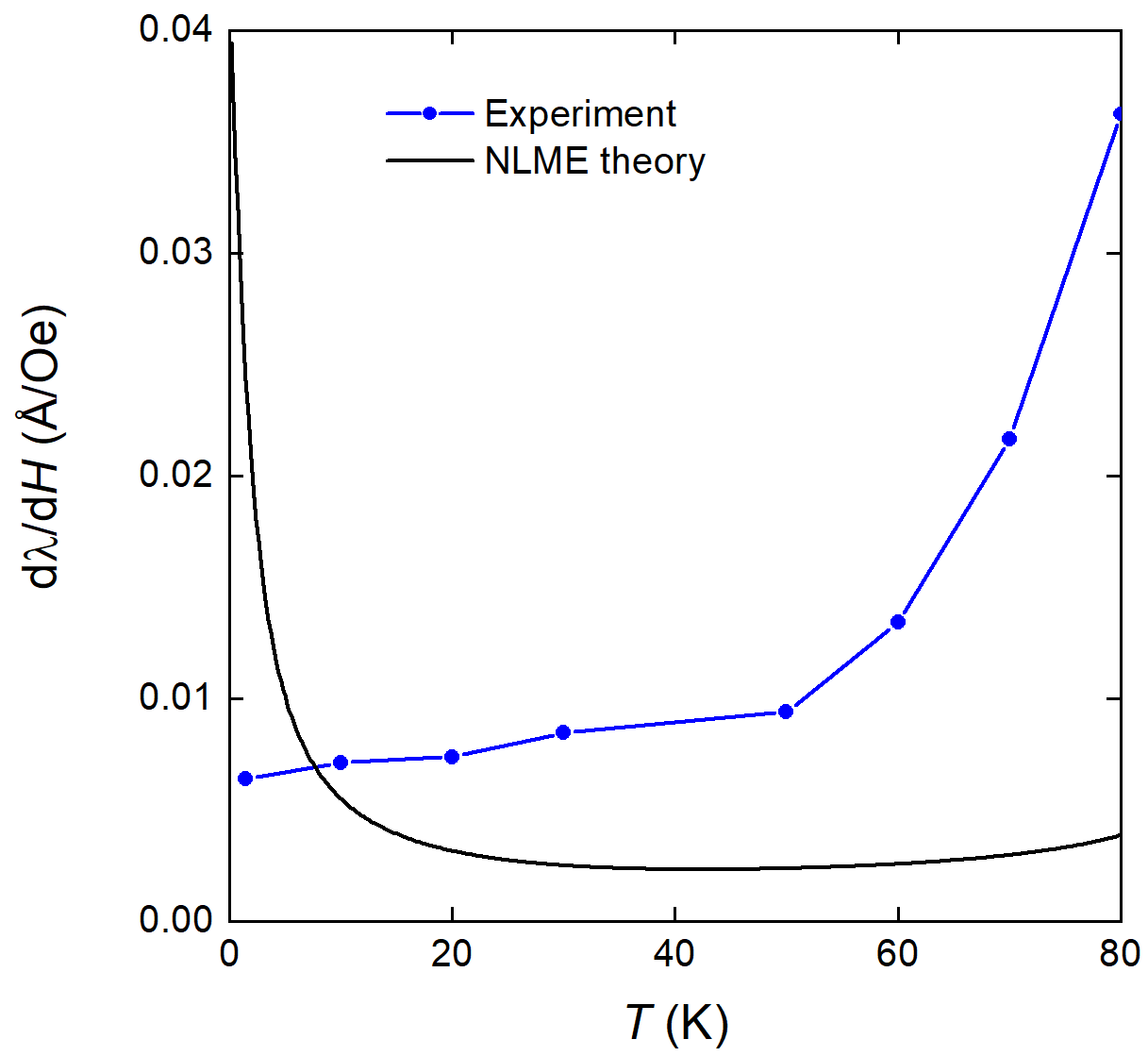}
	\caption{(Color online) Temperature dependence of $d\lambda/dH$ for Y123 as in Fig.\ 8.  The NLME theory line is the calculated change $\Delta\lambda(H)/H$, with $H=300$\,Oe, and assuming that $H_0=2.5$\,T and $\lambda(0)=1400$\,\AA, following the calculation procedure in Refs.\ \cite{Xu95,wilcox}.}
	\label{fig9}
\end{figure}

\noindent Treating the vortex as a particle in a harmonic pinning well subject to an oscillating Lorentz force, the Campbell depth is given by, 
\begin{equation}
	\lambda^2_{\rm{Campbell}} = \frac{B\phi_{0}}{\mu_{0} \alpha_{p}}
\end{equation}
where $\phi_{0}$ is the flux quantum and $\alpha_{p}$ is the effective force constant of the pinning well. A recent, more comprehensive treatment of the Campbell depth may be found in ref. \cite{willa}.   For small fields, the Campbell depth also leads to a linear magnetic field dependence to the measured penetration depth.  However, with increasing temperature, vortex pinning weakens, $\alpha_{p}$ decreases and the Campbell contribution to the penetration depth $increases$, precisely the opposite of the nonlinear Meissner effect.  As shown in figure \ref{fig9}, an increase of $d\lambda /dH$ with temperature is what we observed particularly closer to $T_c$, thus pointing to vortex motion as the cause.  We reluctantly published a null result for the nonlinear Meissner effect in Y123 \cite{carrington} as did Bidnosti  \textit{et. al.} \cite{bidnosti} who used a low frequency mutual inductance bridge method to look for the effect.   Previously, Sridhar et. al. \cite{sridhar} had used a TDR to study $\lambda$ in Y123 and concluded that the field dependence was consistent with pair-breaking in an $s$-wave pairing state.   Later, Maeda et. al. \textit{et. al.} \cite{maeda} used a TDR to observe a linear field dependence to $\lambda$ in Bi$_2$Sr$_2$CaCu$_2$O$_y$.  Although they claimed to have observed the nonlinear Meissner effect, $\lambda(T)$ in their samples had a quadratic (rather than linear) temperature dependence, indicating inadequate purity.  In addition, they found $d\lambda/H$ to be only weakly dependent on temperature, in contradiction with the theory shown in figure 9, and again suggesting that vortex motion was responsible for their observed field dependence.  

\indent It is worth pointing out that the apparent noise in figure \ref{fig8} is actually highly reproducible. Subtracting away the linear trend leaves an oscillation with a period of about 0.5\,Oe, almost certainly coming from a modulation by the DC field of the critical current in weak links somewhere in the sample.  In other superconductors we have observed such oscillations with periods of only a few mOe.  This places an upper limit on the RF field amplitude since the oscillations would not be observable unless the RF field were much less than the DC field periodicity of a few mOe. 

Recently Wilcox \textit{et al.} \cite{wilcox} have succeeded in observing the nonlinear Meissner effect in non-copper oxide superconductors, CeCoIn$_5$ and LaFePO. These are also nodal superconductors but have a much lower $T_c$ (2.3\,K and 6\,K respectively) than optimally doped Y123.  As $H_{c1}$ does not depend on $T_c$ but $H_0$ does, it is expected that the field-dependent change in $\lambda$ at the maximum field before vortices enter varies as $1/T_c$.  Therefore the non-linear Meissner effect is much easier to see in these low-$T_c$ materials.  The effect is likely also present in Y123 but is masked by extrinsic effects such as vortices (above) and Andreev bound states as described below.

\section{Surface Andreev Bound States}

\begin{figure}[tbh]
	\centering
	\includegraphics[width = 8.5cm]{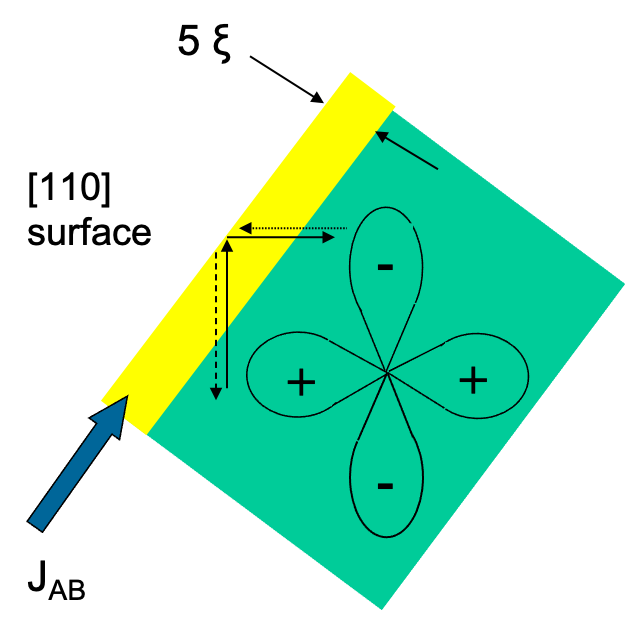}
	\caption{(Color online) $d_{{x^2} - {y^2}}$ gap function in relation to the [110] surfaces of a Y123 sample. Specular scattering at the surface, shown by the arrows, leads to surface Andreev bound states localized near the boundary. $J_{AB}$ is the current from bound states.}
	\label{fig10}
\end{figure}

The $d$-wave order parameter changes sign in $\bm k$-space and it was soon appreciated that this would to lead to current-carrying states localized near particular crystalline faces \cite{hu,tanaka,fogelstrom,barash}. These so-called surface Andreev bound states (ABS) can arise in the geometry shown in figure \ref{fig10} where the $d$-wave gap function is superimposed on a crystal with a [110] facet. The oppositely-signed lobes of the gap function present an effective potential for quasiparticles moving along the trajectories shown, leading to zero energy bound states that live within a few coherence lengths of the [110] surface. These states carry current $J_{AB}$ and their presence leads to a peak in the quasiparticle density of states at zero energy and consequently a $\sim 1/T$ dependence for $\lambda$ that appears below about 10\,K in Y123 crystals. Moreover, the contribution from bound states is predicted to vanish in small magnetic fields as the bound state energy is pushed away from the Fermi energy by the Doppler shift in quasiparticle energies caused by the screening currents.  

\indent Magnetic impurities may also lead to a paramagnetic upturn in $\lambda$ \cite{pcco}. Given the unusual nature of surface Andreev bound states, it was important to distinguish the two. Observations by Carrington \textit{et al.}  \cite{carringtonabs} satisfied several criteria for the $1/T$ upturn. (1) It appears below approximately 10\,K in Y123 (2) Its magnitude is a function of the angle between the actual crystal face and the [110] direction (3) It appears only with induced supercurrents confined to the sample edges (4) It is quenched with a small ($\sim$ 100 Oe) DC magnetic field.  Magnetic impurities would satisfy none of these criteria.

\begin{figure}[tbh]
\centering
	\includegraphics[width = 8.5cm]{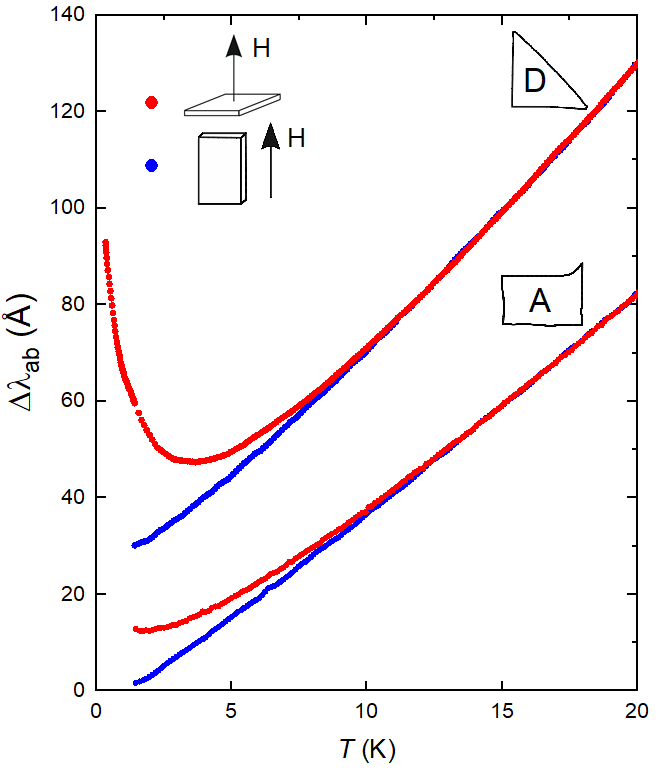}
	\caption{(Color online) Andreev bound states in optimally doped samples of  YBa$_2$Cu$_3$O$_{6+x}$. For RF field $H\|$c (red) an upturn in $\Delta\lambda$ is seen at low $T$ which is much more pronounced for sample D which has is cut to expose more [110] surface as compared to sample A as illustrated by the shape profiles of the samples.  For each sample, there is no upturn when $H\|$ab (blue). }
	\label{fig11}
\end{figure}

\indent Figure \ref{fig11} shows data for two crystals with differing amounts of [110] surface and with the RF field in two orientations.   With the field normal to the planes, induced supercurrents circulate around the perimeter of the crystal and penetrate in from the edge on a length scale of $\lambda_{ab}$, the in-plane penetration depth.  This orientation is expected to show the $1/T$ upturn since it is most sensitive to the bound state current $J_{AB}$ on [110] faces.  If the RF field is oriented parallel to the planes, supercurrents flow across the top and bottom faces where there are no bound states so no $1/T$ upturn is expected.  Figure \ref{fig11} shows the result for each RF field orientation in two different crystals.  In each crystal, an upturn was observed only for the predicted RF field orientation and the size of the upturn scaled as predicted with the amount of [110] surface present.  Paramagnetic upturns in $\lambda$ below $16 K$ were also reported in thin films of Y123 by Walter \textit{et. al.} \cite{walter} Those authors used heavy ion irradiation to generate tracks with varying amounts of [110] surface.  The size of the upturn scaled with the estimated amount of [110] surface] as expected for surface Andreev bound states.  They did not report measurements of the magnetic field dependence of $\lambda$.  

\begin{figure}[tbh]
	\centering
	\includegraphics[width = 8.5cm]{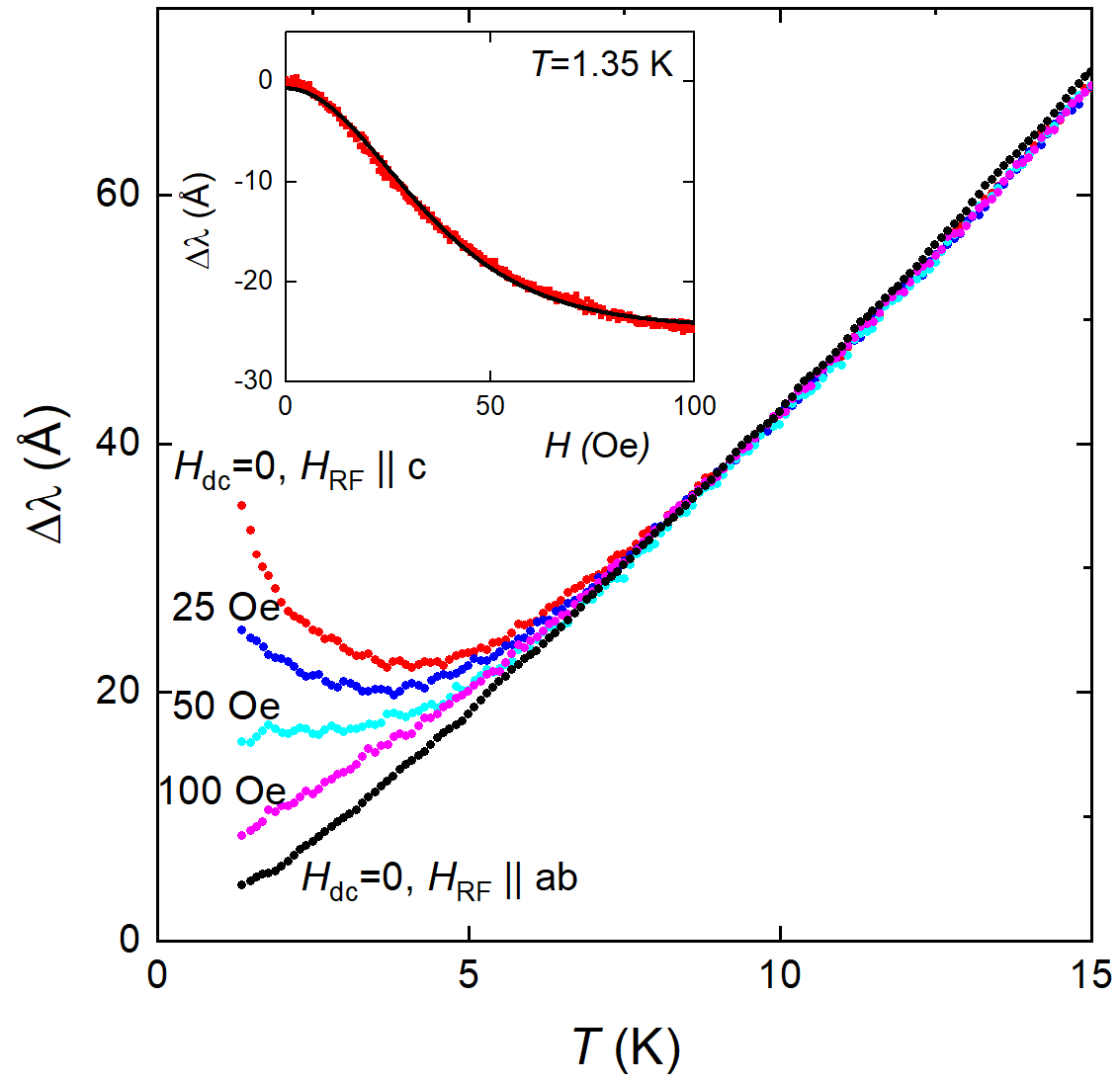}
	\caption{(Color online) Andreev bound state contribution quenched with field. A small dc magnetic field suppresses the upturn in $\lambda(T)$ measured for $H\|$c (sample D, Fig.\ 11).  For $H=100$\,Oe the $H\|$c data is almost the same as  the $H\|$ab.  The inset shows the change in $\lambda$ versus field at a fixed temperature $T=1.34$\,K (the background response measured at $T$=25\,K has been subtracted. The black solid line is a fit to Eq.\ \ref{absH}.  Note the RF fields and dc fields are always parallel. }
	\label{fig12}
\end{figure}

\indent Figure \ref{fig12} shows how the upturn in $\lambda(T)$ is quenched in a dc field of 100 Oe, which is remarkably small.  Were the upturn coming from magnetic impurities, the penetration depth would be enhanced by a factor of $(1+ \chi ({\mu B}/{k_{B}T}))^ {1/2}$ where $\chi$ is the magnetic susceptibility.  Assuming an impurity magnetic moment $\mu = \mu_{B}$, $T = 1 K$ and spin $1/2$, a field of 3 Tesla would be required to reduce $\chi$ by 95 percent.   The much smaller field scale arising from Andreev bound states can be understood from a simple model in which bound states contribute a delta function peak to the density of states at an energy given by the Doppler shift in equation \ref{doppler} ($\delta E=ev_FA$).  This leads to a field-dependent upturn given by,
\begin{equation}
\frac{\Delta \lambda(T,H)}{\lambda(0)} = \frac{\beta}{4T} \cosh^{-2}\left(\frac{\mu_0 e \lambda Hv_F}{2k_BT}\right).
 \label{absH}
\end{equation}

Here $v_F$ is the Fermi velocity and $\beta$ is a function of the angle between the surface normal and the [110] direction \cite{beta}.  The inset shows a fit to Eq.\ \ref{absH}.  A more accurate theory \cite{barash} includes the necessary averaging over the Fermi surface and the effect of finite quasiparticle lifetime (scattering).  Fitting to that theory gives results which are similar to Eq.\ \ref{absH}, but with a higher value of $v_F$.   In fact, $v_F$ is essentially the only free parameter and was determined to be $v_F=1.2\times 10^5$ms$^{-1}$, similar to other estimates. Bound states do not occur unless the gap function changes sign so these observations were strong evidence for the unconventional, $d_{x ^2-y ^2}$ pairing state first proposed for the copper oxide superconductors.   Tsai and Hirschfeld \cite {tsai} later pointed out that both the 1/T upturn and its suppression by low magnetic fields could arise from sufficiently isolated, unitary-limit impurity scatterers in a d-wave superconductor.  Their mechanism does not predict the dependence on crystal shape ([110] surface) that we observed, but it may account for earlier observations of a low temperature upturn in $\lambda$ for Zn-doped Y123 \cite{bonn2}.

\section{Multiband superconductors}

Penetration depth measurements in the high temperature superconductors focused on the gap function in materials with simple Fermi surfaces.  In the majority of superconductors, pairing occurs on multiple Fermi surface sheets but the scattering of electrons between sheets effectively homogenizes the gap function. However, in materials where the Fermi sheets have distinct orbital character, interband scattering is strongly suppressed resulting in distinct gap functions for each sheet, i.e., multiband superconductivity.  We present two examples.  The first is MgB$_2$, a phonon-mediated superconductor with $T_{c} = 39$~K.  The second is CaKFe$_{4}$As$_{4}$ ($T_c=35$~K) a member of the large class known as iron-based superconductors \cite{hosono} in which the pairing appears to be mediated by magnetic interactions \cite{mazin,hirschfeld2011}. 

\begin{figure}[tbh]
	\centering
	\includegraphics[width = 8.5cm]{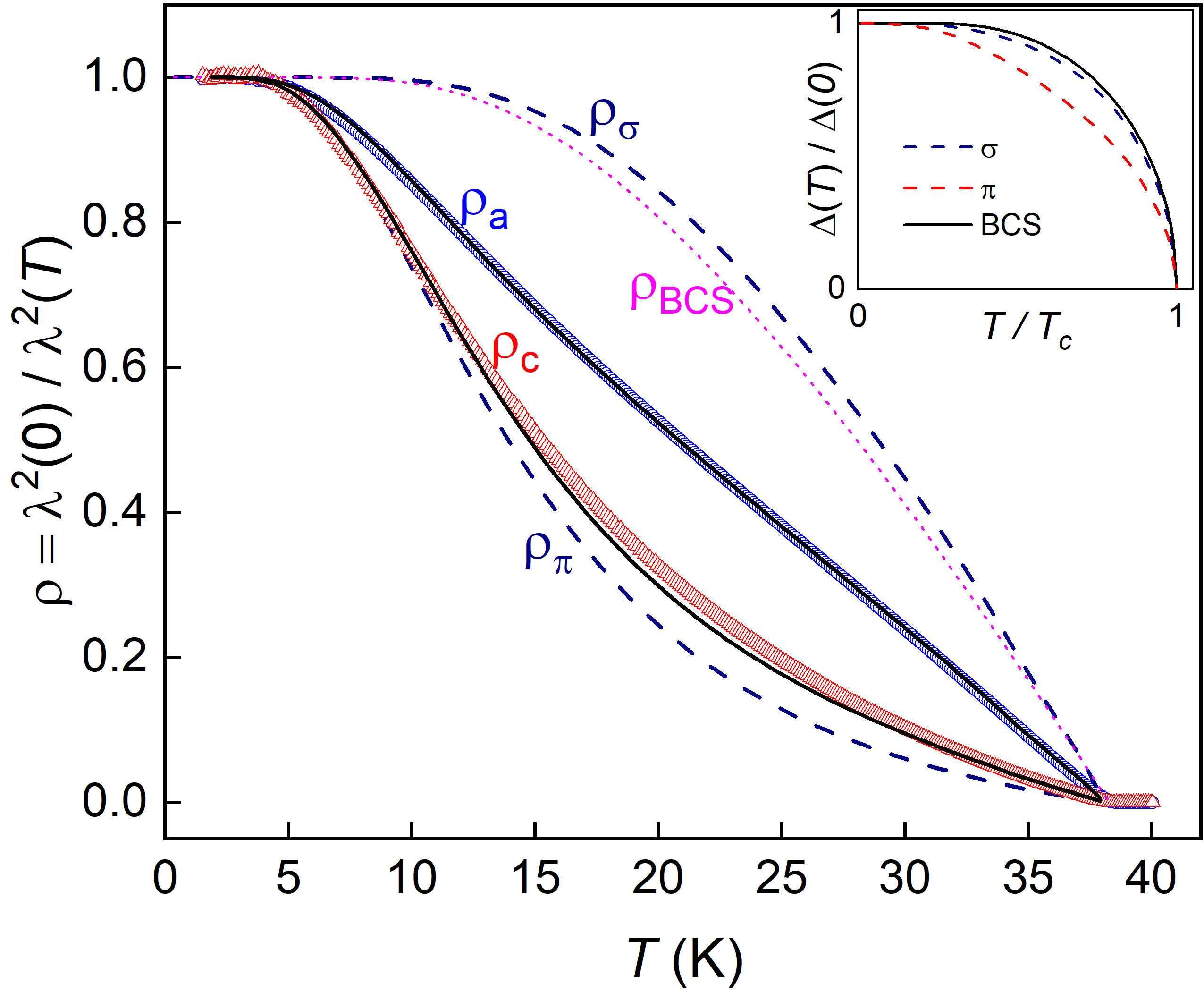}
	\caption{(Color online) Normalised superfluid density $\rho_S$ versus $T$ for MgB$_{2}$. $a,c$ are crystalline axes while ${\sigma},{\pi}$ refer to Fermi surface sheets.  Experimental data is shown by the symbols, whereas the solid lines show the fits to the data.  The individual contributions of the $\sigma$ and $\pi$\ sheets are shown by the dashed lines and the isotropic BCS response by the dotted line. The inset shows the normalised temperature dependence for the gap on each sheet compared to the isotropic BCS prediction with $\Delta(0) = 1.76 {\ }k_{B}T_c$ \cite{fletcher}.}  
	\label{fig13}
\end{figure}

\indent While measurements of $\Delta \lambda$ for $T \ll T_c$ are necessary for detecting nodes and sign changes in the order parameter, measurements over the entire temperature range up to $T_{c}$ are required to identify multiple gaps.  Over this full range, as figure \ref{fig6} shows, the penetration depth looks essentially flat while the superfluid density is far more revealing.  Figure \ref{fig13} shows the results for MgB$_2$\cite{fletcher}.  Superfluid densities $\rho_a$ and $\rho_c$ correspond to screening currents along the $a$ and $c$ axes of the crystal.  Pairing in MgB$_2$ occurs on different Fermi surface sheets (${\sigma}, {\pi}$), each with its own superfluid density ($\rho_{\sigma}$, $\rho_{\pi}$) and gap function (${\Delta_{\sigma}}$, ${\Delta_{\pi}}$).   The temperature dependence of each gap is show in the inset.  The fits in figure \ref{fig13} were produced using a weighted sum,  

\begin{equation}
	\rho_{a,c} = x_{a,c} \rho_{\pi}(T,\Delta_{\pi}) + (1-x_{a,c})\rho_{\sigma}(T,\Delta_{\sigma})
	\label{twogapeq}
\end{equation}

\noindent with $\Delta_{\sigma}(0)${\ }={\ }75\,K  and $\Delta_{\pi}(0)$ = 29\,K and $\Delta(T)$ for each band calculated using a 2-band Eliashberg model. The temperature dependence of $\Delta_\sigma$ is close the the isotropic BCS model but for $\Delta_\pi$ there are significant differences. A fully self-consistent analysis of the experimental data in MgB$_2$ can be found Ref.\cite{Kim2019} and is outlined below.  

\begin{figure}[tbh]
	\centering \includegraphics[width=8.5cm]{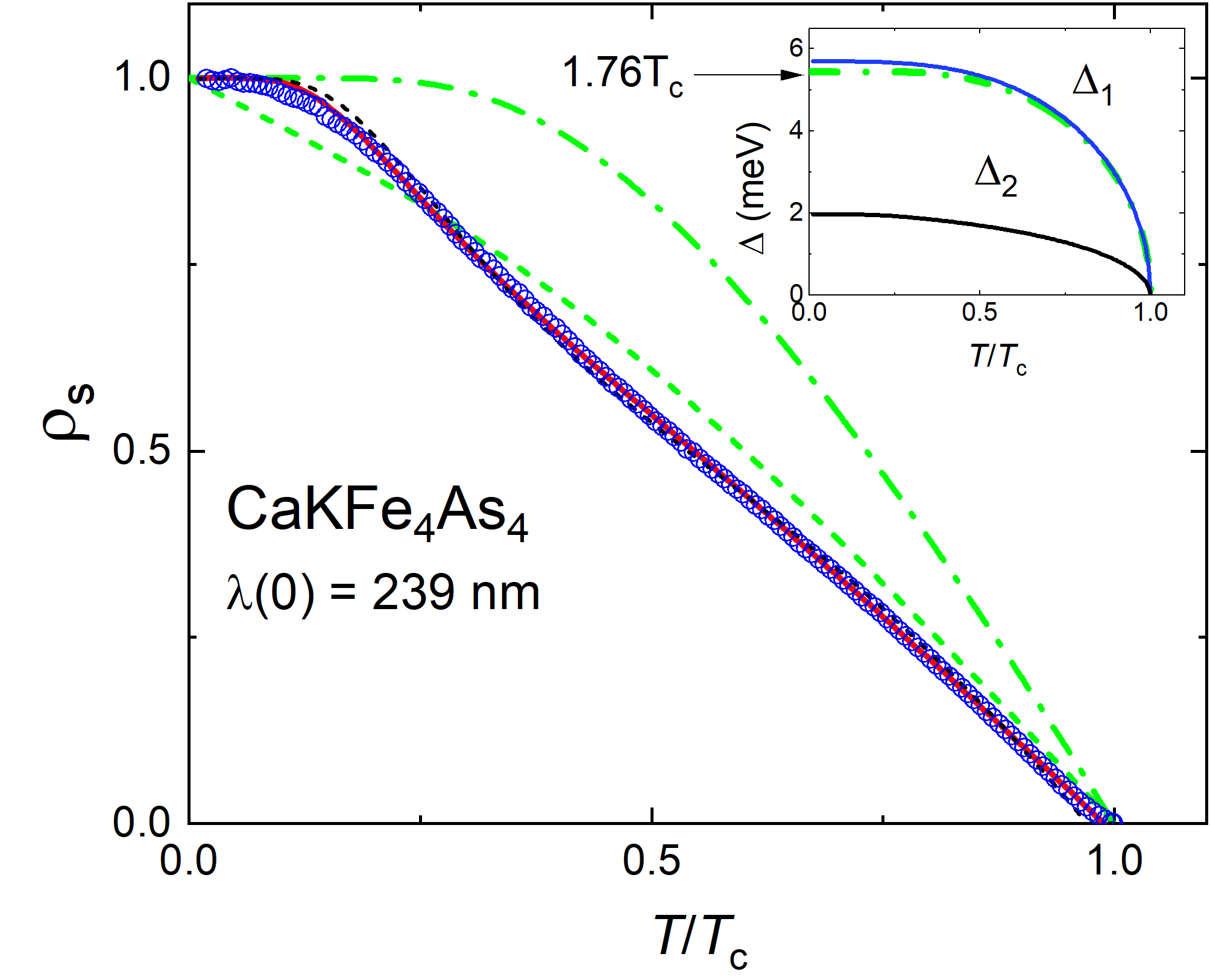} \caption{Superfluid density in CaKFe$_{4}$As$_{4}$.  Solid line is a fit to the self-consistent $\gamma-$model of two-band superconductivity.  The solid green line represents single band
		isotropic weak-coupling s-wave and the dashed green line represents a single band
		d-wave model. (Inset) Gap magnitudes in each band and the weak-coupling single band s-wave gap for comparison. }
	\label{fig14}
\end{figure}

\indent Figure \ref{fig14} shows the superfluid density for CaKFe$_{4}$As$_{4}$, a stoichiometric iron-based superconductor  \cite{Cho2017}.  To generate $\rho_S$ from $\Delta \lambda(T)$ the value of $\lambda(0)$ was taken from the new technique of optical magnetometry using nitrogen vacancy (NV) centers in diamond \cite{Hc1PRA2019}.  
The solid line is a fit to a self-consistent $\gamma-$model of two-band superconductivity involving both in-band and interband pairing interactions as well as two intraband and one interband scattering rate \cite{Kogan2009,prozorov2011}.  For comparison, we also show the superfluid density for a single band $s$-wave (green solid line) and $d$-wave (green dashed line) gap functions.  Both differ significantly from the two-band model, which provides an excellent fit.  The inset shows the gap for each band ($\Delta_1$, $\Delta_2$) calculated self-consistently using a two-band generalization of Eq.\ref{selfcon} \cite{Kogan2009}.  Although the temperature dependence of $\Delta_1$ and $\Delta_2$ are reasonably close to the standard BCS form, one gap is always larger and the other is always smaller than the weak-coupling BCS value. 

\indent An idealized material with only in-band pair interactions would exhibit separate superconducting transitions.  In real materials an interaction always exists and there is just one transition but the shape of $\rho_s(T)$ retains some remnant of the idealized case, sometimes leading to the positive curvature shown by $\rho_{c}$ in figure 13. However, this alone cannot be considered proof of multiband superconductivity and other measurements are needed.  These two examples show that precise penetration depth measurements over the full temperature range, coupled with sophisticated analysis, can now reveal the details of multiband superconductivity. 

\section{Impurity scattering}

Real samples contain impurities, which can substantially alter the temperature dependence of $\lambda$.  While phenomena such as nonlinear Meissner effect and Andreev bound states require the highest purity samples, the deliberate addition of impurities has historically been an important probe of superconductivity.  Anderson's theorem \cite{anderson} states that non-magnetic impurities do not change $T_{c}$ in an isotropic superconductor.  When they {\it do} lower $T_{c}$ a more complicated pairing state is likely to be involved. Impurities may also alter the power law ${\Delta}{\lambda}\sim {T ^n}$ for the temperature dependence of the penetration depth.  The tunnel diode method permits us to measure $n$ with remarkable precision and therefore discriminate among different possible gap functions. Quantitative predictions for the effect of impurities require more advanced mathematical techniques so we will state only the final results \cite{Xu95,prozorov2011,bang,vorontsov}. 

\begin{figure}[tbh]
	\centering
	\includegraphics[width = 8.5cm]{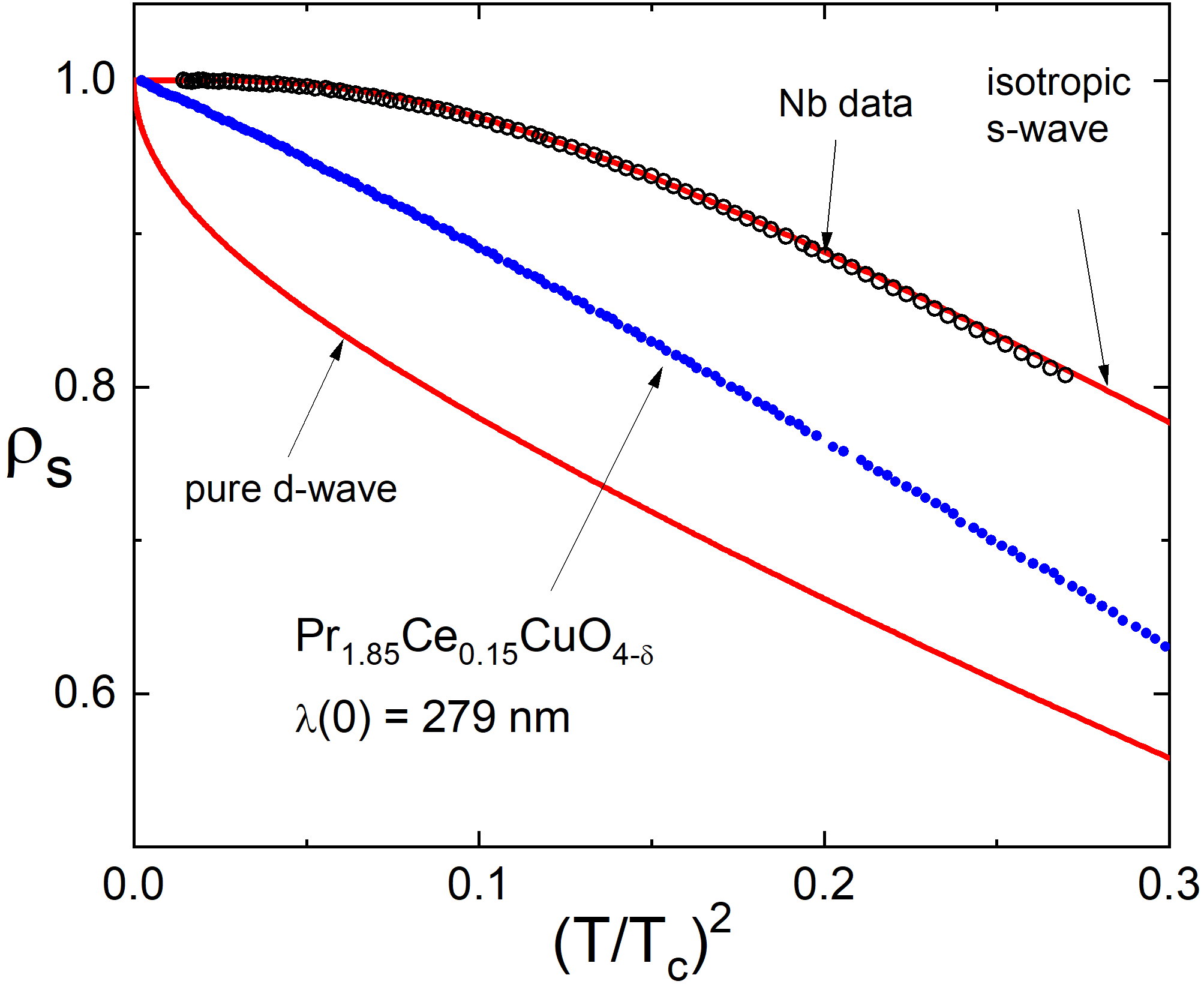}
	\caption{(Color online) $\rho_{S}$ versus $(T/T_{c}) ^2$ for Pr$_{1.85}$Ce$_{0.15}$CuO$_{4-\delta}$ compared with $s-$wave behavior for Nb, taken with the same apparatus.  The quadratic power law is characteristic of a superconductor with line nodes in the presence of strong potential scattering.}  
	\label{fig15}
\end{figure}

\indent Early measurements in the copper oxides showed a quadratic power law ${\Delta}\lambda\left(T\right)\sim T ^2$. This was later shown by Hirschfeld and Goldenfeld to result from non-magnetic impurity scattering in the unitary (strong scattering) limit that generates quasiparticle states at $E$ = 0 without a concomitant sizeable reduction in $T_c$ in a $d$-wave superconductor \cite{hirschfeld1993}. Their empirical formula, valid in the domain $0 \leq T < 0.4 T_c$,

\begin{equation}
	{\Delta}{\lambda}{\sim }\frac{T ^2}{T+T ^*}
\end{equation}

\noindent includes a crossover temperature  ${T ^*}$ related to the density and strength of scatterers.  For $T\gg {T ^*}$, ${\Delta}{\lambda}\sim T$  and for $T < {T ^*}$, ${\Delta}{\lambda}\sim {T ^2}$.   Figure \ref{fig15} shows the superfluid density for the electron-doped copper oxide Pr$_{1.85}$Ce$_{0.15}$CuO$_{4-\delta}$ ($T_{c} = 26$\, K)  \cite{pcco}.  For this material $T < {T ^*}$ in the low temperature region shown.  The $n$ = 2 power law is clearly distinct from the exponential variation seen in Nb and was taken as early evidence for $d$-wave pairing in the electron-doped copper oxides \cite{pcco}.

\begin{figure}[tbh]
	\centering
	\includegraphics[width = 8.5cm]{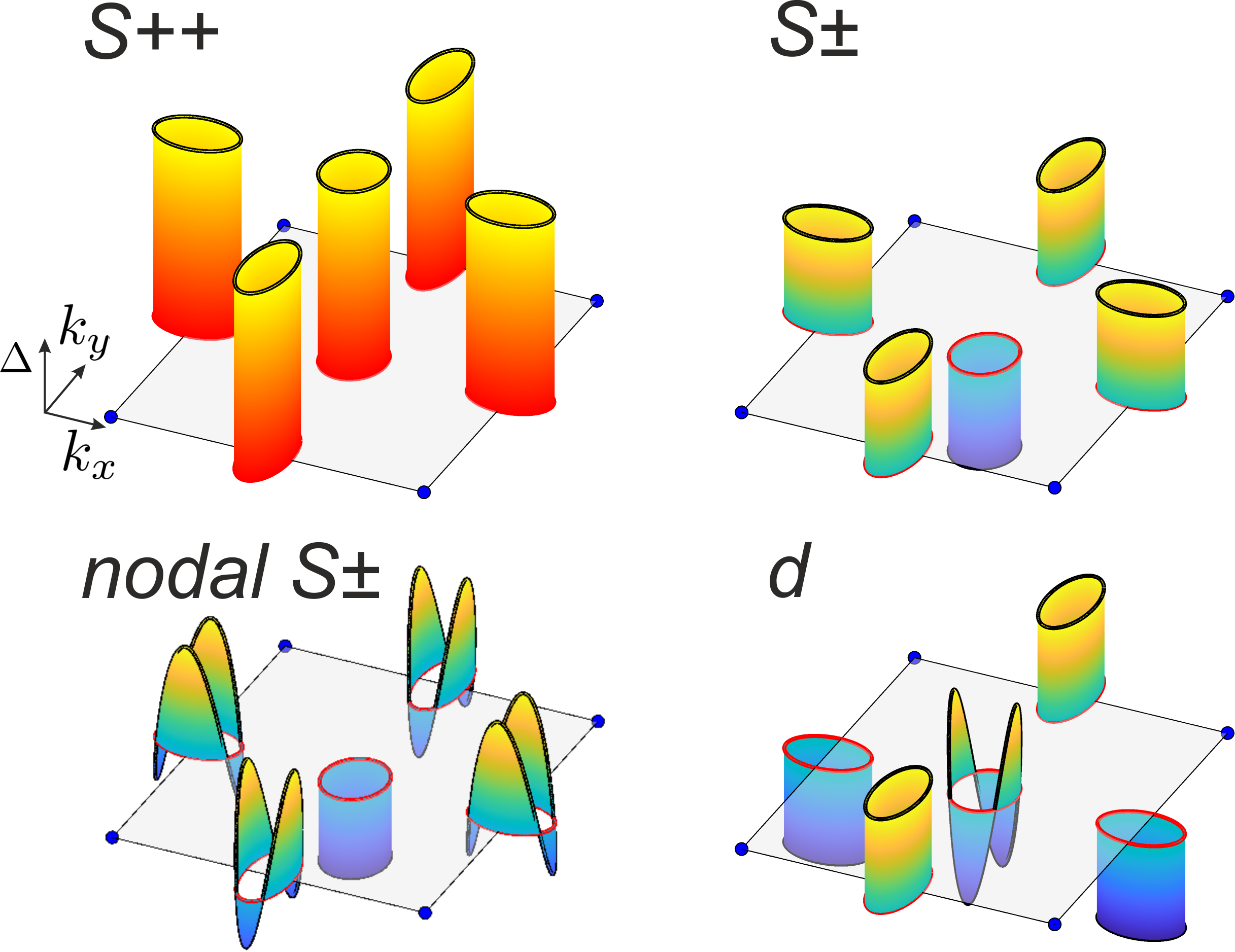}
	\caption{(Color online) Four possible gap functions for iron-based superconductors}
	\label{fig16}
\end{figure}

\indent Penetration depth measurements with added impurities have been particularly useful in revealing the pairing state of the iron-based superconductors.  Since these are multiband superconductors, different gap magnitudes, nodes and sign changes are all possible.  Figure \ref{fig16} illustrates four possible pairing states in the iron-based materials \cite{hirschfeld2011}. In each case, the central surface is hole-like while the four adjacent ones are electron-like.  The leading candidate for most iron-based superconductors is the $s_{\pm}$ state in which hole and electron surfaces are each fully gapped, albeit with different gap amplitudes and a sign change between them. This state is favored by a repulsive interaction coming from the exchange of spin fluctuations between electron and hole sheets \cite{mazin,hirschfeld2011}.  Since it has 4-fold rotational symmetry, the $s_{\pm}$ state is still conventional in a group-theory sense but the sign change leads to measurable consequences.  Interestingly, while Anderson's theorem remains applicable for in-band scattering, nonmagnetic interband scattering {\it is} pair-breaking if the gap function changes sign between bands \cite{Kogan2016,prozorov2011}.

\begin{figure}[tbh]
	\centering
	\includegraphics[width = 8.5cm]{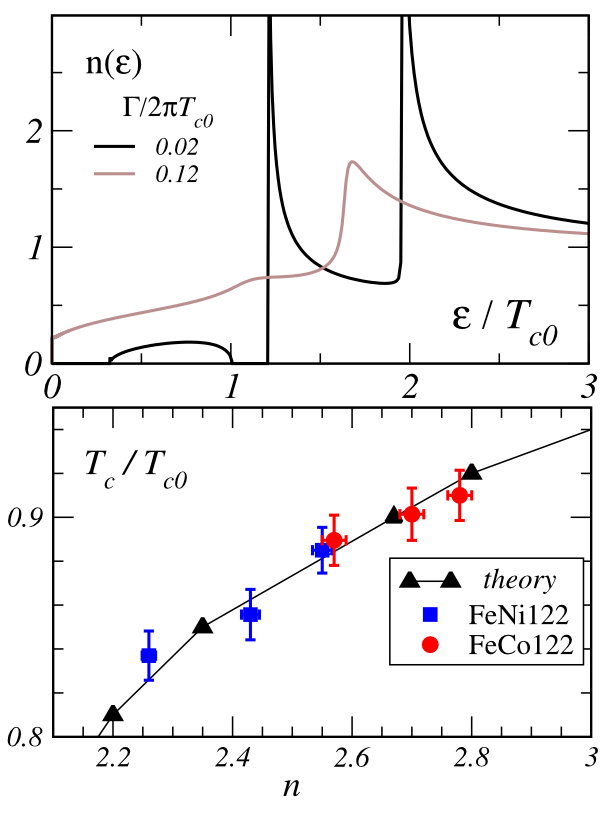}
	\caption{(Color online) (Upper) Calculated density of states versus scaled quasiparticle energy for two different impurity concentrations. At the lower concentration two sharp gap-like features and a band of midgap states are apparent \cite {Kim2010}.  At higher concentration, states appear at zero energy. (Lower) Transition temperature versus exponent $n$ for samples with differing impurity concentration (an implicit variable).}
	\label{fig17}
\end{figure}

Measurements on clean samples cannot easily distinguish $s_{++}$ from $s_{\pm}$. In each case, $\Delta \lambda (T)$ shows an exponential dependence at low temperatures characterized by the smaller of two gaps.   However $s_{++}$ and $s_{\pm}$ superconductors respond differently to the addition of non-magnetic scatterers.  At low temperatures, ${\Delta}{\lambda}$ in an $s_{++}$ superconductor continues to show exponential dependence but in an $s_{\pm}$ superconductor it is predicted to evolve from exponential to $T ^2$  with increased scattering \cite{bang,vorontsov}.  
At our current level of sensitivity, ${\Delta}{\lambda} \sim {T ^{4}}$ is nearly indistinguishable from exponential behavior so the prediction for $s_{\pm}$ is effectively a {\it decrease} of $n$ toward $n$ = 2 with increased scattering.  This contrasts with the case for a $d$-wave superconductor with line nodes, in which more scattering leads to an {\it increase} of $n$ from 1 to 2.  The peculiar situation for $s_{\pm}$ comes from the generation of new quasiparticle states below the minimum energy gap with increased scattering.   The top panel of Figure \ref{fig17} shows the calculated density of states in an $s_{\pm}$ superconductor for two levels of impurity scattering \cite{Kim2010}.  For weak scattering, two sharp gap features are apparent along with a small midband density of states.  Increased scattering  broadens the distribution and generates new states at $E$ = 0 leading to ${\Delta}{\lambda}\sim {T ^2}$. 

\indent  Experiments of this type are sometimes complicated by changes in carrier density when impurities are added.   Columnar defects produced by heavy ion irradiation offer one way around this problem.  This approach was used to vary the density of defects in the electron-doped pnictides Ba(Fe$_{1-x}$Co$_{x}$)As$_{2}$ and Ba(Fe$_{1-x}$Ni$_{x}$)As$_{2}$, as shown in the lower panel of Figure \ref{fig17}. Here $T_c$ is plotted versus the exponent $n$ with the defect density as an implicit variable.  Both $T_{c}$ and $n$ vary exactly as predicted from the density of states calculated for an $s_{\pm}$ order parameter\cite{Kim2010}.  

While the numerical details depend on specifics of the Fermi surface and scattering rates, the {\it evolution} of the power law from high to low and the concomitant change in $T_c$ is a qualitative result that provides some of the strongest evidence for $s_{\pm}$ pairing. Impurity experiments like this may also be useful in identifying topological superconductivity in materials such as the Dirac semimetal, PdTe$_{2}$ \cite{PdTe2eirr2020}.

\section{Absolute penetration depth and quantum criticality}

\begin{figure}[tbh]
	\centering
	\includegraphics[width = 8.5cm]{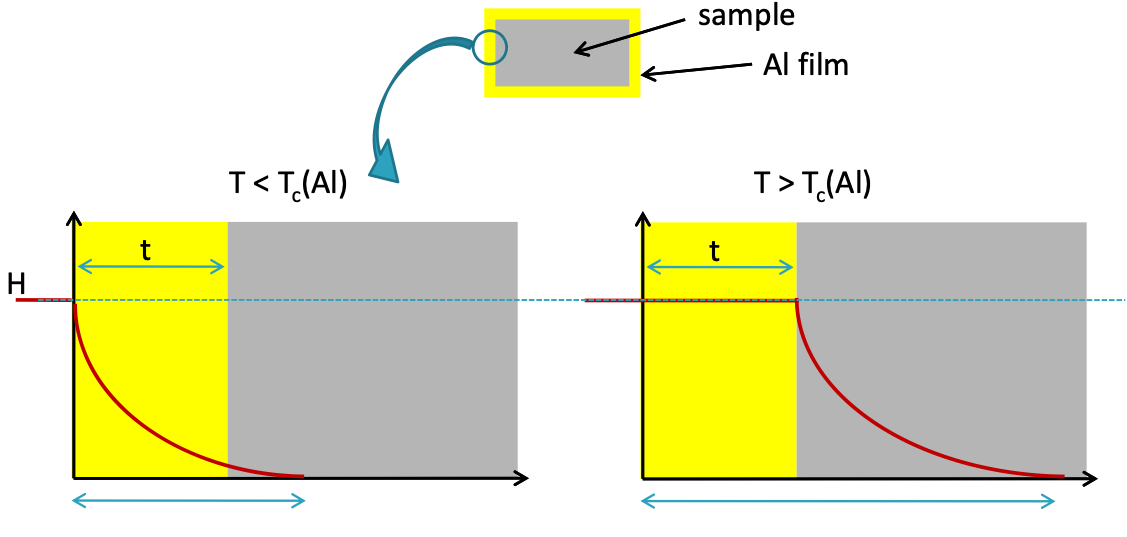}
	\caption{(Color online) Al coating method showing field penetration below and above $T_{c}(Al)$.}
	\label{fig18}
\end{figure}

Although the tunnel diode resonator can easily measure changes in $\lambda$ with very high precision, absolute measurements are considerably more difficult.  Nonetheless, the sensitivity to changes in resonator frequency can be exploited using the scheme shown in Figure \ref{fig18} \cite{Alcoating}.  Here, the sample under study is uniformly sputter-coated with aluminum to a depth of typically 1000\,\AA.  Since the skin depth of Al at 10\,MHz is much larger than  1000\AA, the film is invisible to the RF field when $T > T_{c}(\rm{Al})$ and the effective penetration depth is $\lambda_{eff} = t+{\lambda}$.  This case is shown on the right panel.  As the sample cools below $T_{c}(\rm{Al})$ the film becomes superconducting and the RF field penetrates the composite superconductor shown in the left panel. Treating each superconductor in the London limit, the effective penetration depth of the composite is given by,  

\begin{equation}
	\lambda_{eff} = {\lambda_{Al}}\frac {{\lambda} + {\lambda_{Al}}{\  } \tanh(t/{\lambda_{Al}} ) } {{\lambda_{Al}} + {\lambda}{\  }{\tanh(t/{\lambda_{Al}} )}}
\end{equation}

\begin{figure}[tbh]
	\centering
	\includegraphics[width = 8.5cm]{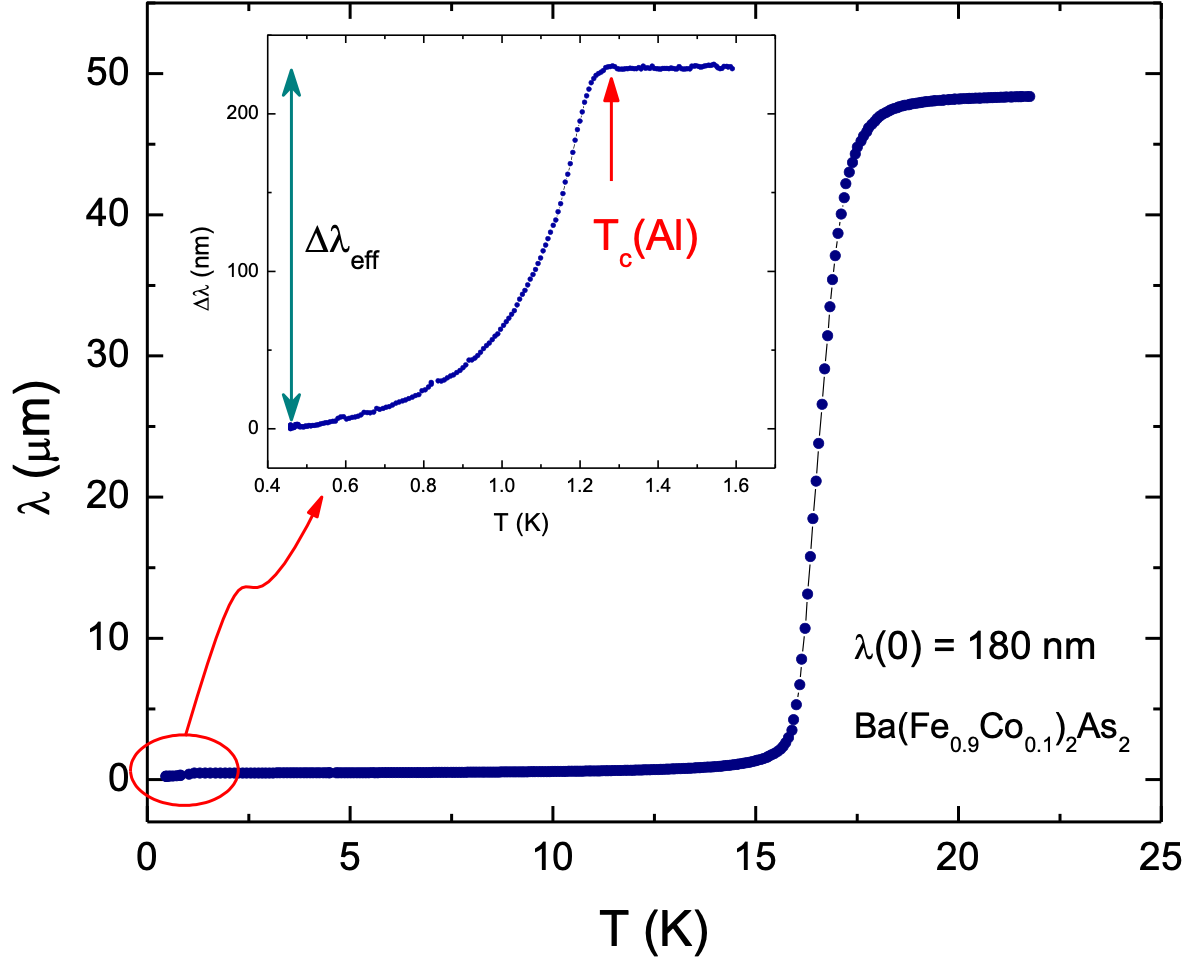}
	\caption{(Color online) Change in penetration depth versus $T$ for Al-coated sample. Inset shows the region in which the Al film becomes superconducting. }
	\label{fig19}
\end{figure}

\noindent By measuring the ${\Delta} \lambda_{eff}$ (shown in the insert) and assuming $\lambda_{\rm{Al}}(0) = 500$\,\AA \cite{tedrow} we can extract $\lambda$ of the underlying superconductor.  Since $T_{c}(Al) = 1.2$\, K is well below that of the underlying superconductor, we essentially measure its zero-temperature penetration depth, $\lambda(0)$.  An example is shown in Figure \ref{fig19} for Ba(Fe$_{0.9}$Co$_{0.1}$)$_2$As$_2$. The inset shows the transition to superconductivity in the Al film.
The inset shows the transition to superconductivity in the Al film. Although there is some uncertainty in $\lambda_{\rm{Al}}(0)$ because of disorder and surface roughness of the film \cite{meservey}, the technique generates values of $\lambda(0)$ for several copper oxide and iron-based superconductors that are in good agreement with other experimental methods.

\begin{figure}[tbh]
	\centering
	\includegraphics[width = 8.5cm]{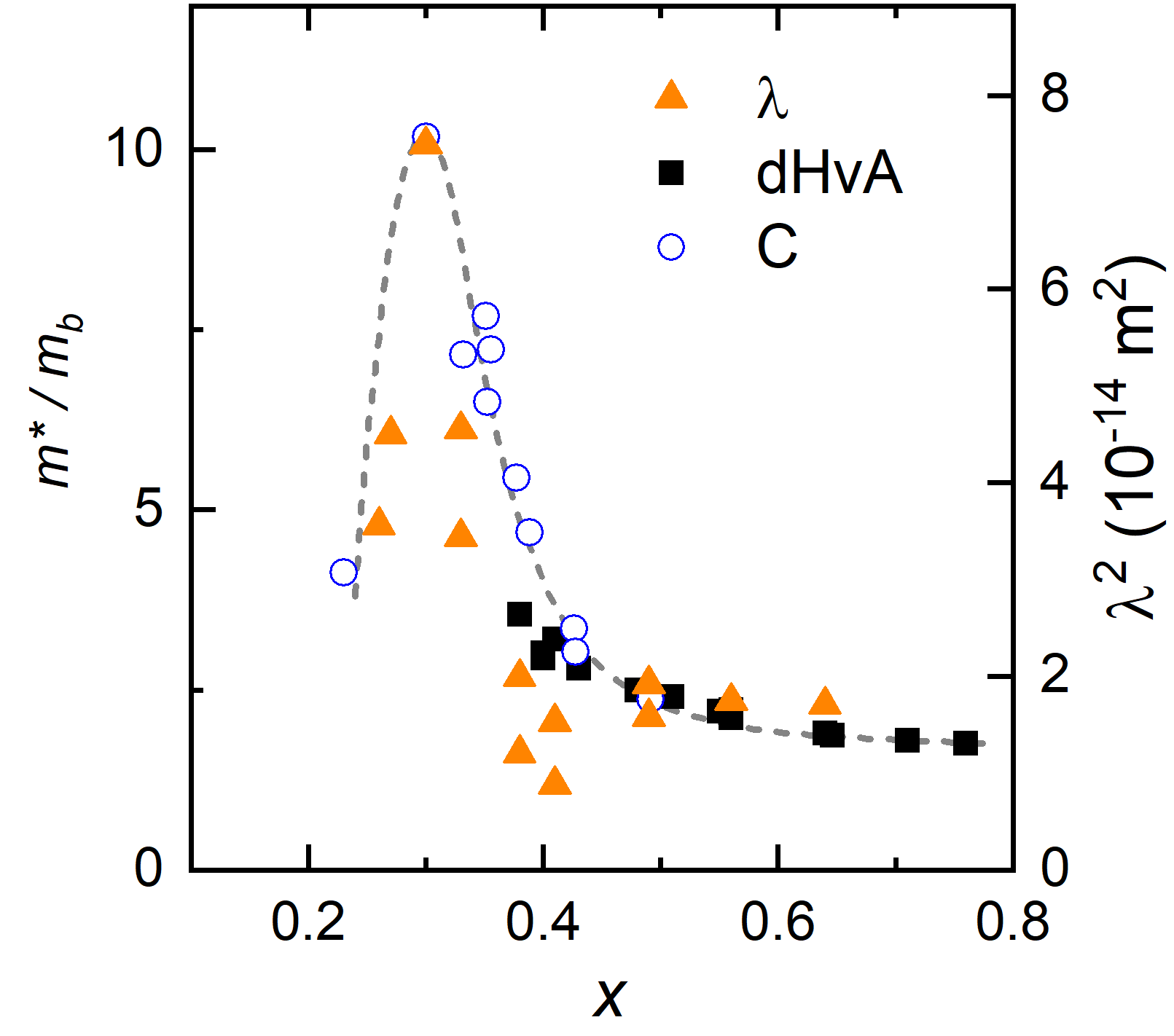}
	\caption{(Color online) A sharp peak in the absolute value of London penetration depth as function of P content $x$ in the iron-based superconductor BaFe$_{2}$(As$_{1-x}$P$_{x}$)$_{2}$ (AsP122) \cite{Hashimoto2012}. The data are compared on an absolute mass enhancement scale with dHvA mass and specific heat data of the same material \cite{walmsley}. As $m\sim\lambda^2n$ includes the weakly $x$ dependent Fermi volume ($n$) derived from dHvA, the right hand $\lambda^2$ axis is only approximate (exact for $x=0.3$).}
	\label{fig20}
\end{figure}

In a simple picture (Eq. 12) $\lambda(0)^2$ is proportional to the ratio of the normal state parameters $n/m$, where $n$ is the volume of the Fermi surface.   In the BaFe$_{2}$(As$_{1-x}$P$_{x}$)$_{2}$ (BaAsP122) system quantum oscillation measurements have shown that $m^*$ varies markedly with $x$, peaking at the composition $x^*=0.3$, which corresponds to a quantum critical point (QCP) of the antiferromagnetic phase \cite{Hashimoto2012,walmsley}.  Measurements of $\lambda$ using the Al plating method also showed a peak at $x^*$ in which the mass enhancement derived from $\lambda(0)$ was found to be in good agreement with those derived both from quantum oscillations and specific heat.  This data is shown in figure \ref{fig20} \cite{Hashimoto2012,walmsley}).  This peak in mass is important evidence that $T_c$ is boosted by the presence of the QCP.   There is some theoretical uncertainty about how mass renormalization $m^*$ enters the expression for $\lambda$. For a Galilean invariant Fermi-liquid the mass renormalization is cancelled out by backflow at $T=0$ (\cite{leggett}), so $m$ in Eq,\ 12 is the bare band mass.  However, it is unclear to what extent this applies to a solid which is not Galilean invariant and the presence of electron and hole pockets may also break the backflow cancellation \cite{Levchenko,nomoto}. The BaAsP122 system is the clearest system to date where this has been tested. 

Later these measurements were extended to Ba(Fe$_{1-x}$Co$_{x}$)$_{2}$As$_{2}$          (BaCo122), where the phase diagram is traversed by electron doping with Co, as opposed to the isoelectric P/As substitution in BaAsP122. This charge doping introduces considerable disorder so in this case quantum oscillations could not be observed. The BaCo122 data show a similar peak in $\lambda$ at the QCP (figure 21) but the higher scatter of the data and the strong increase in $\lambda$ inside the region of the phase diagram where superconductivity coexists with antiferromagnetism makes the peak more difficult to discern \cite{Alcoating} Also, it is important to emphasize that in order to observe a sharp QCP-related peak, several samples of finely spaced compositions are required. The usual case for such compounds is a mismatch between the nominal and actual composition, making such experiments very challenging.

Within conventional theory, a peak in $\lambda(0)$ should cause a corresponding decrease in the lower critical field $H_{c1}$.   However, in BaAsP122 the opposite was found at the QCP  \cite{putzke}. This increase of $H_{c1}$ was attributed to an anomalous increase in the vortex core energy at the QCP \cite{putzke}.  Measurements of $H_{c1}$ in BaCo122 showed the opposite behavior: a clear decrease of  $H_{c1}$ at the QCP \cite{Joshi2020}.  This dichotomy calls for further measurements of $\lambda(0)$ and $H_{c1}$ by different methods on clean systems and those where disorder is added in a controlled way.

\begin{figure}[tbh]
	\centering
	\includegraphics[width = 8.5cm]{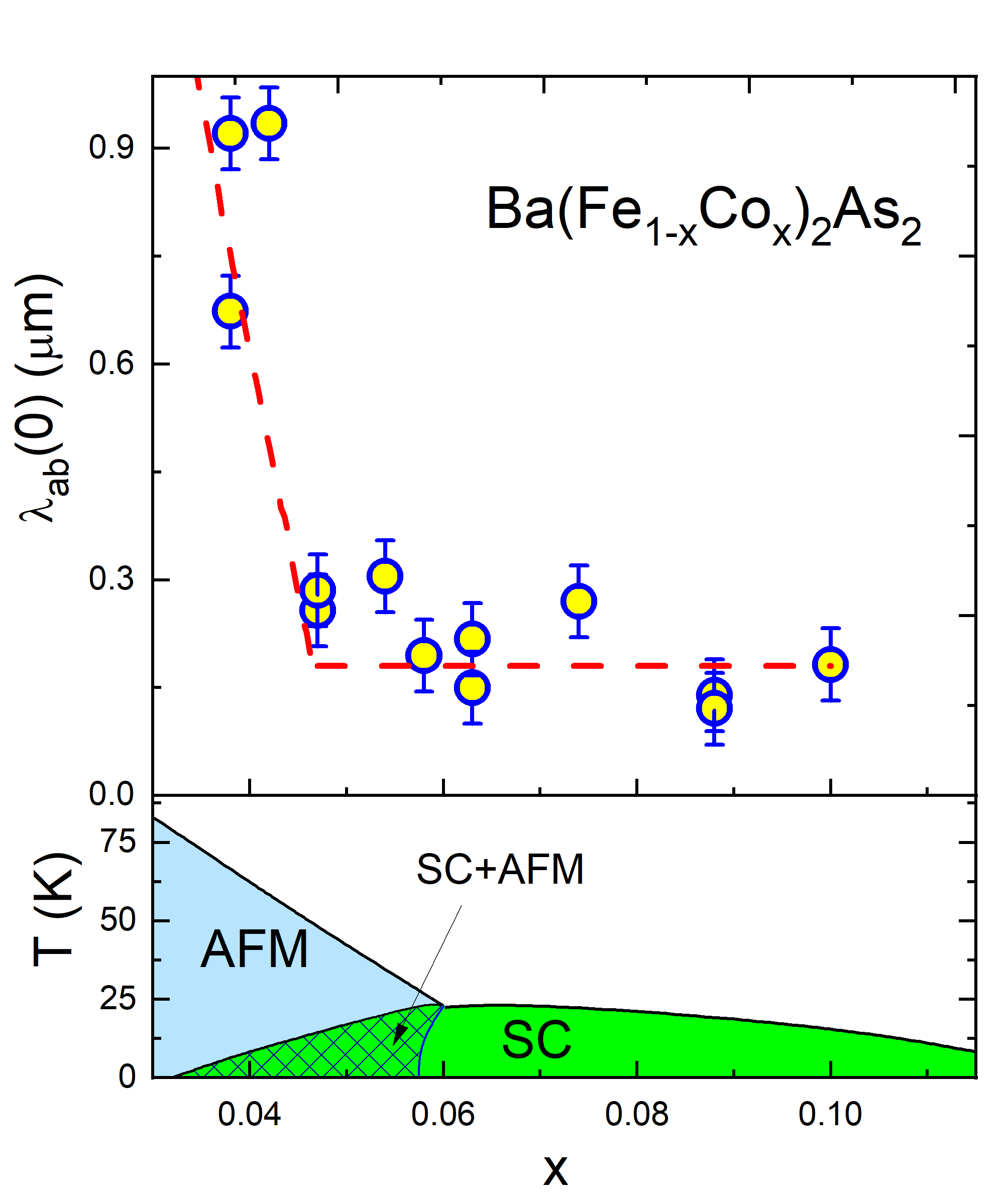}
	\caption{(Color online) (Upper) $\lambda (0)$ for different Co concentrations in Ba(Fe$_{1-x}$Co$_{x}$)$_2$As$_2$. (Lower) Coexistence of itinerant antiferromagnetism (AFM) and superconductivity (SC) under the superconducting dome. }
	\label{fig21}
\end{figure}

\section{Concluding remarks}

This paper presents some highlights from our research using tunnel diode resonators.  There is every reason to believe that this technique will continue to be refined and used to explore new forms of superconductivity.  Our approach was inspired by a Cornell low temperature group seminar given many years earlier.  One prevailing philosophy in that group was the importance of building a unique kind of apparatus with very high sensitivity.  To those of us who were graduate students, the discovery of superfluidity in $^3$He by Osheroff, Richardson and Lee brought that lesson home in a particularly forceful way.  In effect, a Nobel prize hinged on the appearance of a tiny glitch in the $^3$He pressurization curve.  Of course, you had to be cold enough to see it!  A later unexpected jump in the frequency of a torsion oscillator lead to the identification of the Kosterlitz-Thouless transition in $^4$He films by David Bishop and John Reppy. This was a seminal discovery in the physics of topological matter.  RWG is grateful to Dave Lee, John Reppy, Bob Richardson and all the other inhabitants of H-corridor for their insights, camraderie and inspiration during those fantastic years.

\begin{acknowledgements}
	R.P. is supported by the U.S. Department of Energy (DOE), Office of Science, Basic Energy Sciences, Materials Science and Engineering Division. Ames Laboratory is operated for the U.S. DOE by Iowa State University under contract DE-AC02-07CH11358. AC is supported by UK EPSRC grant EP/R011141/1.  RWG was supported by DOE award  DE-AC0298CH1088 (Center for Emergent Superconductivity) and National Science Foundation award numbers NSF-DMR91-20000 and NSF-DMR-05-03882.
\end{acknowledgements}

%
\section*{Conflict of interest}
The authors declare that they have no conflict of interest.


\begin{thebibliography}{}
\bibitem{beta}  $\beta = \frac {\hbar v_{F} g(\theta)}{6 k_{B}\lambda(0)} $ where $g(\theta)$ = $ \left|\left|cos ^{3}(\theta) \right| - \left|sin ^{3}(\theta)\right| \right|$ and $\theta$ is the angle between the surface normal and the [110] axis.
\end{thebibliography}


\end{document}